\documentclass{aa}

\DeclareMathAlphabet{\altmathcal}{OMS}{cmsy}{m}{n}
\usepackage{mathptmx}

\usepackage[T1]{fontenc}
\usepackage{txfonts}

\usepackage{graphicx}	% Including figure files
\usepackage{amsmath}	% Advanced maths commands
\usepackage{amssymb}	% Extra maths symbols
\usepackage{gensymb}
\usepackage{siunitx}
\usepackage{xcolor}     % Extra maths symbols
\usepackage{cancel}
\usepackage{multirow}
\usepackage{mathrsfs}
\usepackage{ulem}
\usepackage[hidelinks,colorlinks=true,linkcolor=blue,citecolor=blue]{hyperref}
\renewcommand{\vec}[1]{\mathbf{#1}}
\renewcommand{\tens}[1]{\mathsf{#1}}

\newcommand{\pd}[2]{\frac{\partial #1}{\partial #2} }
\newcommand{\der}[2]{\frac{\mathrm{d} #1}{\mathrm{d} #2} }

\newcommand{\DS}{\displaystyle}

\newcommand{\change}[1]{{#1}}

\DeclareMathOperator{\sign}{sign}

\DeclareSIUnit{\rad}{rad}
\DeclareSIUnit\dyne{dyn}

\newcommand{\km}{\;\mathrm{km}}
\newcommand{\s}{\;\mathrm{s}}
\newcommand{\ms}{\;\mathrm{ms}}
\newcommand{\erg}{\;\mathrm{erg}}
\newcommand{\gauss}{\;\mathrm{G}}
\newcommand{\gcm}{\;\mathrm{g\;cm^{-3}}}
\newcommand{\dyncm}{\;\mathrm{dyn\;cm^{-2}}}

\begin{document} 

   \title{Magnetic dissipation in short gamma-ray burst jets}

   \subtitle{I. Resistive relativistic MHD evolution in a model environment}

   \author{Giancarlo Mattia    \inst{1}\fnmsep\thanks{mattia@fi.infn.it}
      \and Luca Del Zanna \inst{2,1,3}
      \and Andrea Pavan     \inst{4,5}
      \and Riccardo Ciolfi    \inst{4,5}
          }

   \institute{
INFN , Sezione di Firenze, Via G. Sansone 1, I-50019 Sesto Fiorentino (FI), Italy 
%\\ \email{mattia@fi.infn.it}
\and        Dipartimento di Fisica e Astronomia, Universit\`a di Firenze, Via G. Sansone 1, I-50019 Sesto Fiorentino (FI), Italy
\and        INAF, Osservatorio Astrofisico di Arcetri, Largo E. Fermi 5, I-50125 Firenze, Italy
\and        INAF, Osservatorio Astronomico di Padova, Vicolo dell'Osservatorio 5, I-35122 Padova, Italy
\and        INFN, Sezione di Padova, Via F. Marzolo 8, I-35131 Padova, Italy
             }

\date{Received XXX; accepted YYY}

% \abstract{}{}{}{}{} 
% 5 {} token are mandatory
 
  \abstract
   {}
   {Short gamma-ray bursts originate when relativistic jets emerge from the remnants of binary neutron star mergers, as observed in the first multimessenger event GW170817 - GRB 170817A, in coincidence with a gravitational wave signal. Both the jet and the remnant are believed to be magnetized, and the presence of magnetic fields is known to influence the jet propagation across the surrounding post-merger environment. In the magnetic interplay between the jet and the environment itself, effects due to a finite plasma conductivity may be important, especially in the first phases of the jet propagation. We aim to investigate such effects, from jet launching \change{towards} its final breakout from the post-merger environment.}
   {2D axisymmetric and full 3D resistive relativistic MHD simulations, employing spherical coordinates with spatial radial stretching, are performed with the PLUTO numerical code.  Different models for physical resistivity, which must be small but still above the numerical one (producing unwanted smearing of structures in any \textit{ideal} MHD code) are considered and compared. Stiff terms in the current density are treated with \textit{IMplicit-EXplicit Runge Kutta} algorithms for time-stepping. A Synge-like gas (Taub equation of state) is also considered. All simulations are performed by using an axisymmetric analytical model for \change{both} the jet propagation environment \change{and the jet injection}; we leave the case of jet propagation in a realistic environment (i.e. imported from an actual binary neutron star merger simulation) to a later study.
   %, and a more realistic one, imported from fully 3D numerical relativity simulations of a binary neutron star merger.
   }
   {As expected, \change{no qualitative differences are detected due to the effect of a finite conductivity}, but significant quantitative differences in the jet structure and induced turbulence are clearly seen in 2D axisymmetric simulations, and we also compare different resistivity models. Regions with a resistive electric field parallel to the magnetic field form and non-thermal particle acceleration may be enhanced there. The level of dissipated Ohmic power is also dependent on the various recipes for resistivity. Most of the differences arise before the 
   %\change{breakout from the inner weakly magnetized environment, whereas once the jet enters the external environment region}
   \change{breakout from the inner environment, whereas once the jet enters the external weakly magnetized environment region}
   these differences are preserved during further propagation despite the lower grid refinement. Finally, we show and discuss the 3D evolution of the jet within the same environment, in order to highlight the emergence of non-axisymmetric features.}
   {}


   \maketitle

%--------------------------------------------------------------------
\section{Introduction}
%--------------------------------------------------------------------

The gravitational wave (GW) detection of the binary neutron star (BNS) merger event GW170817, followed after $1.74$\,s by the detection of the short gamma-ray burst (sGRB) GRB\,170817A and later by follow-up electromagnetic observations across the whole spectrum, was the first multimessenger event involving a GW source \citep{Abbott_etal_2017c, Abbott_etal_2017a, Abbott_etal_2017b}.
This breakthrough event confirmed in particular that BNS mergers can produce a powerful relativistic jet associated with a sGRB \citep{Mooley_etal_2018, Ghirlanda_etal_2019}. 
However, various aspects of the post-merger dynamics remain elusive, including the precise nature of the merger remnant acting as the central engine for launching the jet.

In general, prompt black hole (BH) formation will occur for high-mass BNS mergers, while lower-mass systems may lead to hyper-massive neutron stars (HMNSs) in a temporary equilibrium before collapsing to a BH, or even to stable NSs. 
In the GW170817 case, models tend to favor the formation of a remnant surviving the collapse for about 0.5--1 second (e.g., \citealt{Margalit_Metzger_2017}), leaving the question of whether the jet was launched before or after BH formation.
In the former case, the jet could be powered by a central magnetar-like object (e.g., \citealt{Ciolfi_etal_2017, Ciolfi_etal_2019}) in which the field has been enormously amplified by shear during the merging process itself (e.g., \citealt{Palenzuela_etal_2022} and refs.~therein), via the magneto-rotational instability (e.g., \citealt{Moesta_etal_2020}), or via internal mean-field dynamo action (e.g., \citealt{Franceschetti_etal_2020, DelZanna_etal_2022, kiuchi_etal_2024}).
This scenario is however matter of debate, with serious difficulties pointed out in the last few years \citep{Ciolfi_2020a}. The alternative BH central engine scenario seems to find stronger support in numerical relativity simulations, where the possibility of forming at least a magnetically dominated and mildly relativistic incipient jet has been demonstrated (e.g., \citealt{Ruiz_etal_2016}).

GRB 170817A was also a peculiar sGRB, the first observed 15--30 degrees away from the jet propagation axis and well outside the jet core opening angle (of only a few degrees; \citealt{Mooley_etal_2018, Ghirlanda_etal_2019}). This strongly boosted the investigation on how incipient jets propagate through post-merger environments and acquire their specific angular structure (e.g., \citealt{
Lazzati_etal_2018,
Xie_etal_2018,
Geng_etal_2019,
Kathirgamaraju_etal_2019,
Gottlieb_etal_2020,
Nathanail_etal_2020,
Gottlieb_etal_2021,
Pavan_etal_2021,
Lazzati_etal_2021,
Murguia-Berthier_etal_2021,
Nathanail_etal_2021,
Urrutia_etal_2021,
Gottlieb_etal_2022b,
Pavan_etal_2023,
Urrutia_etal_2023,
Hamidani_Kunihito_2023}).

To accurately model the above propagation and the emerging jet properties, a `realistic'  environment (i.e.~directly imported from the outcome of a BNS merger simulation) has been recently shown to represent a necessary ingredient \citep{Pavan_etal_2021, Lazzati_etal_2021, Pavan_etal_2023}.
Moreover, magnetic fields in the environment need to be included (along with those within the jet) as they can have a significant impact on the post-breakout angular structure and the final energetics of the jet-cocoon system (\citealt{Pavan_etal_2023, Garcia-garcia_etal_2023}).
In this case, however, the complex interaction between the jet magnetic field and that of the surrounding environment may suffer from excessive smoothing due to numerical viscosity (or more properly, resistivity) in MHD simulations, especially in three dimensions where the resolution cannot be too high.

\change{Such ``spurious'' dissipation, originated by truncation errors and highly susceptible to the grid resolution and the numerical algorithms employed, can strongly impact our understanding of the evolution of astrophysical environments (see, e.g., \citealt{Puzzoni_etal_2021,Puzzoni_etal_2022}).
On the other hand, numerical simulations of relativistic plasmas encompassing a physical resistivity model that dominates over the numerical diffusion (see, e.g., \citealt{Palenzuela_2013, Dionysopoulou_etal_2015, Qian_etal_2018, Inda-Koide_etal_2019, Vourellis_etal_2019}) showed the importance of a dissipative model on the simulation outcome and how a physical resistivity is not only crucial for the reproducibility of the results (which, for instance, are not affected by the algorithms employed) but also on the consistency of the different physical processes in the act.}

Therefore, the goal of the present series of two papers is to investigate, for the first time, the magnetic dissipation and in general the effect of a finite plasma conductivity in two-dimensional (2D) axisymmetric and three-dimensional (3D) relativistic MHD (RMHD) simulations of sGRB jets propagating through a BNS merger environment, along the lines of our previous work on generic astrophysical jets in a uniform medium \citep{Mattia_etal_2023}. The main result of that work is the finding that the fine current sheet structures and the level of turbulence are affected by the value of the plasma conductivity, for which both uniform and variable models (based either on a tracer or on the local Alfvén speed) are tested. Regions with a non-ideal electric field parallel to the local magnetic field form, and plasmoids inside current sheets are visible if the resistivity is low and the resolution is high enough. The magnetic dissipated power is also computed.

In the present paper (Paper I), the BNS merger environment is provided by an analytical axisymmetric model, describing a magnetized wind in radial expansion, with a homologous velocity profile, surrounded by a static external atmosphere (recipes in \citealt{Pavan_etal_2021}). The magnetic field of the wind, in particular, is provided by a dipole of initial strength $\sim 4\times 10^{13}\gauss$, superimposed to the hydro structure. Into such a wind, an incipient sGRB jet, magnetized and uniformly rotating, is continuously injected from the inner boundary at a given radius, covering an opening angle of $10\degree$, with initial Lorentz factor 3 and total luminosity of about $10^{52}\erg/\mathrm{s}$. The evolution is followed for $0.5\s$ after launching, when the jet has eventually broken out from the expanding environment, up to a distance of about $10^5\km$. A detailed comparison between 2D axisymmetric simulations and 3D ones, assuming the same setup for both the environment and the incipient jet is provided here, while we leave to a forthcoming Paper II the presentation and discussion of the fully 3D case with a realistic (non-axisymmetric) environment, imported following the guidelines in \citet{Pavan_etal_2023}.

The resistive relativistic MHD (RRMHD) simulations are performed using spherical coordinates with the PLUTO code \citep{PLUTO_2007}. We employ state-of-the-art methods for resistivity in shock-capturing evolution schemes and we assume a Synge-like equation of state, more appropriate than the usual ideal gas condition for a relativistic plasma (for references and details see \citealt{Mattia_etal_2023}).

The present Paper I is structured as follows: in section \ref{Sec:numerical} we describe the system of equations solved, the numerical methods employed, and the setup for the initial and boundary conditions; section \ref{Sec:results2D} is devoted to the presentation and discussion of the results of axisymmetric 2.5D simulations; in section \ref{Sec:results3D} we compare those results against full 3D runs, using the same axisymmetric initial setup. Conclusions are drawn in section \ref{Sec:conclusions}. Appendix \ref{App:ideal} shows a comparison with an ideal run.

%--------------------------------------------------------------------
\section{Equations and numerical setup}
\label{Sec:numerical}
%--------------------------------------------------------------------

In this section, we describe the equations adopted (i.e. the equations of RRMHD), as well as the numerical setup employed throughout the entire paper.

%-----------------------------------------------------
\subsection{Equations of RRMHD}

The system of RRMHD equations consists of a set of hyperbolic partial differential equations with non-zero source terms \citep{Komissarov_2007, DelZanna_etal_2016, Mignone_etal_2019}
\begin{equation}
  \pd{U}{t} = -\nabla\cdot\tens{F}(U) + S_e(U) + S_i(U)
\end{equation}
where $U$ is the vector of conserved variables, $F$ represents the fluxes and $S_e$ and $S_i$ represent, respectively, the standard (explicit) and stiff (implicit) source terms.
By splitting the equations for the different conserved variables we get
\begin{equation}
\begin{array}{l}
\DS\pd{D}{t} + \nabla\cdot(D\vec{v}) = 0 , \\ \noalign{\medskip}
\DS\pd{\vec{m}}{t} + \nabla\cdot [\rho h \gamma^2\vec{v}\vec{v} - \vec{E}\vec{E} - \vec{B}\vec{B} + (p + \frac{E^2 + B^2}{2})\,\tens{I}] = \vec{f} , \\ \noalign{\medskip}
\DS\pd{\cal E}{t} + \nabla\cdot\vec{m} = {\cal P}_\vec{f}, \\ \noalign{\medskip}
\DS\pd{\vec{B}}{t} + \nabla\times\vec{E} = 0 , \\ \noalign{\medskip}
\DS\pd{\vec{E}}{t} - \nabla\times\vec{B} = -\vec{J}, \\ \noalign{\medskip}
\nabla\cdot\vec{B} = 0, \\ \noalign{\medskip}
\nabla\cdot\vec{E} = q
\end{array}
\end{equation}
where we have {\change{set} $c=1$ and we have absorbed a factor $1/\sqrt{4\pi}$ into the definitions of the electromagnetic fields.

The conserved fluid variables $(D,{\vec m}, {\cal E})$ are, respectively, the laboratory frame density, momentum density, total energy density, ${\vec B}$, ${\vec E}$ and $q$ represent the magnetic and electric fields and the electric charge density. The primitive fluid variables $(\rho,{\vec v}, p)$ are, respectively, the rest-mass density, the velocity, and the pressure of the fluid, $h$ is the relativistic enthalpy function and $\gamma$ is the Lorentz factor of the flow. Here $\vec{f}$ is a generic external force per unit volume, and ${\cal P}_\vec{f}$ the corresponding power per unit volume (see the PLUTO user guide\footnote{https://plutocode.ph.unito.it/userguide.pdf} for a general description and Section \ref{sec::eq_gravity} for the particular implementation needed for the present problem).

The conserved fluid variables to be evolved in time are defined in terms of the primitive variables as
\begin{equation}
\begin{array}{lcl}
D & = & \rho\gamma,  \\  \noalign{\medskip}
\vec{m} & = & \rho h\gamma^2\vec{v} + \vec{E}\times\vec{B}, \\ \noalign{\medskip}
{\cal E} & = & \rho h\gamma^2 - p + \DS\frac{E^2 + B^2}{2},
\end{array}
\end{equation}
while, due to the high non-linearity of the RRMHD equation, the conversion from conserved to primitive variables must be performed numerically \citep{Mignone_etal_2019}.

The electric current density ${\vec J}$ appearing in the source terms of the RRMHD equations is defined in terms of the electromagnetic fields as
\begin{equation}
\vec{J} = q\vec{v} + \eta^{-1}\gamma [\vec{E} + \vec{v}\times\vec{B} - (\vec{E}\cdot\vec{v})\vec{v}]
\end{equation}
This expression comes from the simplest possible form of Ohm's law written for the electric field and current density in the frame comoving with the fluid, that is
\begin{equation}\label{eq:ohm}
e^\mu = \eta j^\mu
\end{equation} 
where $\eta$ is the resistivity coefficient, supposed to be a scalar for the sake of simplicity, and where the comoving electromagnetic fields have been split as
\begin{equation}\label{eq:emu+bmu}
\begin{array}{lll}
e^\mu & = & \gamma (\vec{v}\cdot\vec{E}\,, \,\vec{E} + \vec{v}\times\vec{B}) \\ \noalign{\medskip}
b^\mu & = & \gamma (\vec{v}\cdot\vec{B}\,, \,\vec{B} - \vec{v}\times\vec{E})
\end{array}
\end{equation}
The above expression for ${\vec J}$ is finally obtained  from the projection $j^\mu = J^\mu + (J^\nu u_\nu)u^\mu$, where $J^\mu = (q,{\vec J})$ is the four-current \citep{Bucciantini_DelZanna_2013,DelZanna_Bucciantini_2018,Mignone_etal_2024}.

Following \cite{Mattia_etal_2023}, the set of RRMHD equations is closed by assuming the Taub equation of state \citep{Mignone_Mckinney_2007, Mizuno2013}, which is the appropriate one when a relativistic fluid (the jet) interacts with a non-relativistic medium
\begin{equation}
h = \DS\frac{5}{2}\Theta + \sqrt{\DS\frac{9}{4}\Theta^2 + 1}.
\end{equation}
where $\Theta=p/\rho$ is a temperature function.

In addition, the evolution of a passive scalar tracer $\xi$ is also considered, in order to properly disentangle the jet from the ambient medium (the tracer will be initialized only at the injection boundary). By combining the appropriate advection equation for $\xi$ with the continuity equation for $D$, the code actually evolves in time the following additional conservative equation
\begin{equation}
    \pd{(D\xi)}{t} + \nabla\cdot(D\xi\vec{v}) = 0
\end{equation}

%-----------------------------------------------------
\subsection{Numerical Setup}

All the simulations have been performed with the PLUTO code \citep{PLUTO_2007} on a static mesh in spherical coordinates ($r,\theta,\phi$). We refer to the code's user guide for the recipes employed in the code to handle non-cartesian geometries (in particular, the $\phi$ component of the momentum equation is multiplied by $r$). In some of the plots, cylindrical coordinates ($R=r\sin\theta,z=r\cos\theta$) are introduced when a generic meridional plane is displayed.

The numerical algorithms adopted in this paper show a few slight differences from the ones described in \citet{Mattia_etal_2023}.
In particular, we carry out our simulations using the IMplicit-EXplicit Runge-Kutta SSP3 (3,3,2) scheme \citep{Pareschi_Russo_2005, Palenzuela_etal_2009} coupled with a fifth-order (instead of fourth-order) Piecewise Parabolic reconstruction method \citep{Mignone_2014}.
Instead of the Harten, Lax, and van Leer Riemann solver \citep{HLL_1983} adopted in \citet{Mattia_etal_2023}, we chose to employ an adaptation of the FORCE Riemann solver \citep{Mattia_Mignone_2022} where the fast waves are approximated with the speed of light (which does not only represent the maximum speed reachable but also the propagation of the fast waves in the frozen limit, as shown in \citealt{Mignone_etal_2018}).
As in \citet{Mattia_etal_2023}, we preserve the solenoidality of the magnetic field through the divergence cleaning algorithm.
Note that the divergence cleaning method requires an additional evolutionary equation
\begin{equation}
    \DS\pd{\psi}{t} + c_h^2\nabla\cdot\vec{B} = -\DS\frac{c_h^2}{c_p^2}\psi
\end{equation}
which, as in \citep{Mignone_etal_2010_glm}, is solved in two steps:
\begin{itemize}
    \item the advective step is solved within the IMEX framework
    \item the source term is solved by using its analytical solution
    \begin{equation}
        \psi^{\Delta t^n} = \psi^{(0)}\exp\left(-\DS\frac{c_h^2}{c_p^2}\Delta t\right)
    \end{equation}
\end{itemize}
For this simulations we set $c_h = 1$, $c_p^2 = \Delta h/\alpha^2$, $\alpha = 0.01$ and $\Delta h = \min(\Delta r)$.
We point out that the GLM method also impacts the evolutionary equation of the magnetic field as well, which becomes
\begin{equation}
\pd{\vec{B}}{t} + \nabla\times\vec{E} + \nabla\psi = 0
\end{equation}
Conversely, the divergence of the electric field is not explicitly enforced, as in \citet{DelZanna_etal_2016, Tomei_etal_2020, Mattia_etal_2023}.

Due to the high Lorentz factors involved, the implicit IMEX step may become brittle due to the high non-linearity of the RRMHD equations.
In particular, we found two potential issues:
\begin{itemize}
\item as already shown in \citet{Mattia_etal_2023}, if the four-velocity is non-negligible in any of the components, convergence may not be reached due to an endless increase of the Lorentz factor. Therefore, if an iteration during the implicit step becomes higher than $10^3$ (since, as shown in the next subsections, we expect the maximum Lorentz factor achievable by our simulations to be $<300$) we reset the four-velocity to 0 and the other primitive variables to the ones of the previous timestep;
\item The initial guess provided for the implicit step may not always be physical, since it is computed from the explicit step of the IMEX scheme. Such a step, without the implicit update, is still incomplete and therefore the conserved variable resulting from it may not represent a physically valid configuration. If the velocity computed from the intermediate step exceeds the speed of light we set its initial guess to 0.9 (in units of $c$). Such a fix involves only the velocity since all the other fluid variables do not require a guess during the implicit step.
\end{itemize}
The code robustness is additionally enhanced by employing the shock flattening methods present in the PLUTO code (see the code's user guide).

%-----------------------------------------------------
\subsection{Numerical grid}

In this paper, we present two types of simulations.
The first kind is a 2D axisymmetric parameter study where half of the angular domain is explored, namely $\theta \in [0, \pi/2]$ covered by $N_\theta = 256$ grid points, and therefore no counter-jet is shown.
The second kind is a half 3D grid (i.e. with no counter-jet, as for the 2D simulations) rotated by $\pi/2$ so that the jet symmetry axis is on the equatorial plane, in order to fully explore the jet dynamics without the potential numerical issues or artifacts coming from the polar axis singularity.
In 3D simulations the angular directions $\theta \in [0.1,\pi - 0.1]$ and $\phi \in [0,\pi]$ are covered by, respectively, $N_\theta = 480$ and $N_\phi = 512$ grid points in order to preserve the resolution adopted in the 2D axisymmetric simulations.
\begin{figure}
  \centering
  \includegraphics[width=0.49\textwidth]{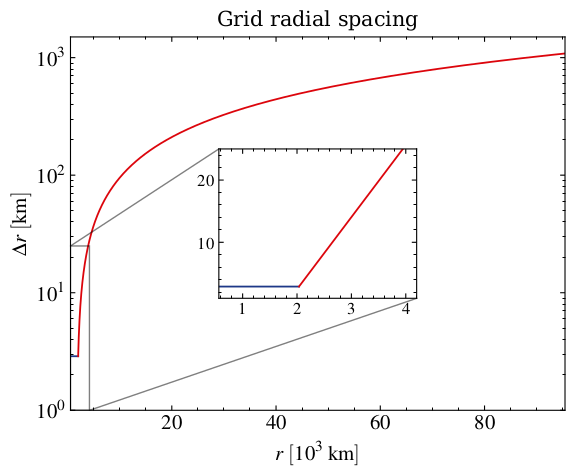}
  \caption{\footnotesize Radial spacing as a function of the distance from the center. The zoomed panel highlights the boundary between the uniform- and the stretched-spaced regions, plotted, respectively, in blue and red for the sake of clarity.
  }
  \label{fig:grid}
\end{figure}

The radial direction is spaced in the same way for all the simulations and, as shown in Fig. \ref{fig:grid}, is designed to capture the dissipative effects closer to the excision radius and to maintain their impact up to the larger scales.
To achieve such a goal we have constructed a uniform grid from $r_{\min} = 562.6\km$ to $r_{\mathrm{bound}} = 2036.3\km$ covered by 512 grid points (yielding a spacing of $\Delta r \approx 3\km$).
On top of the uniform grid, we have constructed a stretched grid up to $r_{\max} = 95439.3\km$, reaching the coarsest grid spacing of $\sim10^3\km$ near the outer boundaries.
The 2D and 3D simulations are run, respectively until $t = 10^5$ and $t = 2\times10^4$ in code units, which corresponds to $t = 0.4926 \s$ and $t = 0.0985\s$.

In the remainder, we will describe both the initial conditions for $t=0$ and the jet's injection recipes for $t>0$. For the sake of simplicity, we will just describe the setup for the 2D axisymmetric case. In the 3D case, the vector fields are tilted by $90\degree$ to arrange the orbital axis orthogonally to the cylindrical z-axis, as anticipated.
For instance, the rotated expression for the magnetic field becomes (the full calculation can be found in Appendix C of \citealt{Pavan_etal_2023})
\begin{equation}
\label{eq:vec_tilt}
\begin{array}{lcl}
\vec{B} & = & B^r \vec{e}_{r'} + \left(B^\theta\sin\psi\sin\phi + B^\phi\DS\frac{\cos\psi}{\sin\theta}\right)\vec{e}_{\theta'} + \\ \noalign{\medskip}
&& \left(-B^\theta\DS\frac{\cos\psi}{\sin\theta} + B^\phi\sin\psi\sin\phi\right)\vec{e}_{\phi'}
\end{array}
\end{equation}
where here the primes indicate the new 3D reference system, with
\begin{equation}
    \psi = -\sign(\cos\theta)\arccos\left(
    \DS\frac{\sign(\cos\phi\sin\theta)}{\sqrt{1 + 
    (\cos\phi\tan\theta)^{-2}}}\right)
\end{equation}
and the same rotation is applied to the velocity.

%-----------------------------------------------------
\subsection{Initial conditions}

Due to the highly complex structure of the post-merger magnetic field in realistic BNS merger simulations, a detailed numerical study of resistive/dissipative processes in such cases is rather challenging. In this paper, for the sake of simplicity, the jet is launched into an idealized analytic environment, similar to the test case presented (along with the more realistic one) in \citet{Pavan_etal_2021}. 
We postpone to a subsequent work (Paper II) the study of resistive effects within the realistic post-merger environment directly imported from the outcome of a BNS merger simulation.

We start by defining a radius $r_{\mathrm{b}} = 7383.2 \km$, which represents the (initial) boundary between the radially expanding, magnetized post-merger environment and the external, unperturbed, unmagnetized static atmosphere.
The density profile is defined as
\begin{equation}
    \rho \propto \left\{
    \begin{array}{lcl}
    \left(\DS\frac{r}{r_{\min}}\right)^{-s} & \qquad & r \leq r_{\mathrm{b}} \\ \noalign{\medskip}
    \left(\DS\frac{r}{r_{\mathrm{b}}}\right)^{-6.5} & \qquad & r \geq r_{\mathrm{b}}
    \end{array}\right.
\end{equation}
with continuity for $r_{\mathrm{b}}$, where $\rho (r_{\min})= 1.056\times10^8\gcm$ and $s = 3.981$. A similar formula applies for the pressure, with $p (r_{\min})= 6.408\times10^{25}\dyncm$ and $s = 3.320$.

The post-merger wind is described by a radial velocity with a homologous increase up to $r_{\mathrm{b}}$
\begin{equation}
    v^r/c = 0.047\left(\DS\frac{r}{r_{\min}}\right) - 0.037
\end{equation}
while we assume $\vec{v} = 0$ in the static atmosphere.
As for the velocity, we assume $\vec{B} = 0$ in the atmosphere, while the post-merger magnetic field is assumed to be a dipole up to $r_{\mathrm{b}}$, with
\begin{equation}
    \begin{array}{lcl}
    B^r = 2B_{\mathrm{d}}\cos\theta\left(\DS\frac{r_{\min}}{r}\right)^3 \\ \noalign{\medskip}
    B^\theta = B_{\mathrm{d}}\sin\theta\left(\DS\frac{r_{\min}}{r}\right)^3 
    \end{array}
\end{equation}
with $B_\phi =0$, where $B_{\mathrm{d}} = 4.013\times10^{13} \gauss$.
Finally, the electric field is initialized everywhere to its ideal value ($\vec{E} = -\vec{v}\times\vec{B})$ (thus null in the static atmosphere), while the passive tracer is set to 0, given that the jet will be injected from the boundary only for $t>0$, as described below.

%\subsubsection{Realistic environment}

%{\RED Explain realistic simulation import B5e15

%Explain the link with the atmosphere}

%-----------------------------------------------------
\subsection{Boundary conditions and jet injection}

Due to the different types of simulations presented in this paper, the boundary conditions are also depending on the configuration adopted.
In the 3D simulation, we adopted, respectively, zero-gradient (outflow) and periodic boundaries for the $\theta'$ and $\phi'$ coordinates.
Conversely, in the 2D simulations, we impose axisymmetric and equatorial symmetric
boundary conditions at $\theta = 0$ and $\theta = \pi/2$, respectively.
The outer radial boundaries are set to outflow in all the simulations, while the inner radial boundaries are defined in order to inject the magnetized jet after the collapse phase ($30\mathrm{\;ms}$ after the NS collapse) into the computational domain. The recipes for the injection are summarized in the following steps (a more complete derivation can be found in \citealt{Pavan_etal_2023}), in which dimensional units are introduced:

\begin{enumerate}

\item we set the jet opening angle $\theta_j =10\degree$, hence the jet is injected where $\theta < \theta_j$, and we also set the tracer $\xi = 1$ there (zero elsewhere);

\item we set the injection parameters as in \citet{Pavan_etal_2023}. In particular, the injection and termination Lorentz factors are set, respectively to $\gamma_j =3$ and $\gamma_\infty = 300$, the ratio between the maximum toroidal and radial magnetic field strength is set to $B_{\rm rat} = 0.5$, the jet angular velocity is set to $\overline{\Omega}\sim10\;\mathrm{rad/s}$, the maximum azimuthal magnetic field is set to $B_{j}= 1.55\times10^{13}\gauss$, the magnetic field half-opening angle is set to $\theta_{j,m} = 4\degree$, and the jet density at the edge is set to $\rho(\theta_j) = 2.203\times10^{3}$ g/cm$^3$;

\item we compute the injection velocity and magnetic field at the initial injection time
\begin{equation}
\begin{array}{lcl}
    \vec{v}_{0} & = & \left(c\sqrt{1 - \gamma_j^{-2}}, \;0,\; \overline{\Omega}r\sin\theta\right) \\ \noalign{\medskip}
    \vec{B}_{0} & = & \left(B_{\rm rat}B_{j}, \; 0,\; 
    \DS\frac{2B_j(\theta/\theta_{j,m})}{1 + (\theta/\theta_{j,m})^2} \right)
\end{array}
\end{equation}
and we also compute $b^2(\theta)$ using the expression for $b^\mu$ in the case of the ideal MHD approximation, where we re-introduce the correct units
\begin{equation}
b^\mu = \frac{1}{\sqrt{4\pi}}\left[\frac{\gamma (\vec{v}\cdot\vec{B})}{c}\,, \,\frac{\vec{B}}{\gamma} + \frac{\gamma (\vec{v}\cdot\vec{B})\vec{v}}{c^2}\right]
\end{equation}
%
%\begin{equation}
%\begin{array}{ll}
%    \rho(\theta) & = \rho(\theta_j)\exp\left[-\DS\int^{\theta}_{\theta'_j} f(\theta')d\theta'\right] + \exp\left[-\DS\int^{\theta}_{\theta'_j} f(\theta')d\theta'\right] \\  \noalign{\medskip}
%    & \times\DS\int^{\theta}_{\theta'_j}\left\{\exp\left[ 
%    \DS\int^{\chi}_{\theta'_j} f(\theta')d\theta' \right]
%    g(\chi)d\chi\right\} \geq b^2/(c^2h_0^*) - 1
%\end{array}
%\end{equation}
\item we compute the jet density by assuming
transverse equilibrium between total pressure gradient, centrifugal force, and magnetic tension
\begin{equation}
\rho(\theta) = [ \, \rho(\theta_j) +  \int^{\theta}_{\theta_j} \exp{(F(\theta'))}g(\theta')d\theta' \,]\,\exp{(-F(\theta))}
\end{equation}
where 
\begin{equation}
%\DS\int^{\theta}_{\theta'_j} f(\theta')d\theta' 
F(\theta)
=-\DS\frac{4h^*_0\gamma^2\overline{\Omega}^2r_{\min}^2}{c^2 ( h^*_0 - 1)}
\DS\frac{\sin^2(\theta) - \sin^2(\theta_j)}{2} - \ln\DS\frac{\sin(\theta)}{\sin(\theta_j)}
\end{equation}
and
\begin{equation}
    g(\theta) = -\DS\frac{4(b^\phi)^2 - b^2 + \sin\theta\DS\frac{db^2}{d\theta}}{c^2 (h_0^* - 1)\sin\theta}
\end{equation}
with $h_0^* = \gamma_\infty/\gamma_0$.
%A lower limit for the density
%
%\begin{equation}
%\rho (\theta) \geq \frac{b^2}{c^2(h_0^* - 1)}
%\end{equation}
%
%is enforced to avoid negative pressure values in the next steps.
Subsequently, the jet's initial two-sided luminosity is computed as
\begin{equation}
    L_j = 4\pi r_{\min}^2\DS\int_0^{\theta_j}\left[
    \rho h^*_0\gamma_j^2c^2 - \left(p + \DS\frac{b^2}{2}\right)
     - (b^0)^2\right]v^r\sin\theta d\theta
\end{equation}
where the initial pressure is computed as
\begin{equation}
\label{eq::pressure_jet}
    p = [\rho c^2(h^*_0 - 1) + b^2]/4.
\end{equation}
In all the simulations the jet parameters yield a luminosity of $L_j = 1.08\times 10^{52}\erg/\mathrm{s}$, \change{which corresponds to an initial jet local magnetization}

\begin{equation}
\label{eq::sigma}
    \sigma = \DS\frac{b^2}{4\pi\rho c^2 h}
\end{equation}
\change{between $\sigma_{\mathrm min} = 0.066$ and $\sigma_{\mathrm max} = 11.52$, depending on the angle $\theta$. Notice that we are using here the fully relativistic definition of the (hot) magnetization parameter, as appropriate for high-speed flows. The corresponding relative magnetic luminosity contribution is}
\begin{equation}
\DS\frac{L_j - L_{j,HD}}{L_j} \sim30\%
\end{equation}
where $L_{j,HD}$ is the jet luminosity without the electromagnetic field contribution.
\change{Therefore, at the injection site, the jet can be considered as strongly magnetized \citep{Gottlieb_etal_2020}.}
Note that this step is computed only at the beginning of the simulation and not at every evolutionary step;

\item we apply the decay of the luminosity, by adopting 
\begin{equation}
\label{eq:decay_vars}
\begin{array}{lcl}
    \gamma(t) & = & \sqrt{1 + (\gamma_jv^r_0e^{-t/(2\tau_d)}/c)^2} 
        \\ \noalign{\medskip}
    v^r(t) & = & \DS\frac{\gamma_jv^r_0e^{-t/(2\tau_d)}}{\gamma(t)} 
        \\ \noalign{\medskip}
    B^\phi(t) & = & B^\phi_0\sqrt{\DS\frac{v^r_0}{v^r(t)}}
    e^{-t/(2\tau_d)}{\gamma(t)} 
        \\ \noalign{\medskip}
    B^r(t) & = & B^r_0\sqrt{\DS\frac{v^r_0}{v^r(t)}}
    e^{-t/(2\tau_d)}{\gamma(t)}
        \\ \noalign{\medskip}
    h^*(t) & = & \DS\frac{h^*_0\gamma_je^{-t/(2\tau_d)}}{\gamma(t)}
\end{array}
\end{equation}
where the quantities with the label $0$ refer to the initial time, using the values specified in steps 3 and 4, while $\tau_d = 300\ms$ represents the temporal decay timescale.
The pressure is computed again from Eq. \ref{eq::pressure_jet} with the time-evolved values of the magnetic field (in both the laboratory and comoving frames) and velocity.

\item we compute the electric field under the ideal approximation $\vec{E} = -\vec{v}\times\vec{B}$.

\end{enumerate}

In the full 3D case we recall that all vectors must be rotated according to the recipes previously mentioned. The condition for the jet's injection simply becomes
\begin{equation}
 r_{\mathrm{cyl}} = r|\cos\theta|\sqrt{1 + (\cos\phi\tan\theta)^2} <= r\sin\theta_j   
\end{equation}

%-----------------------------------------------------
\subsection{Gravity}
\label{sec::eq_gravity}

\change{As shown in the simulations of \citet{Pavan_etal_2021}, in the vicinity of the denser parts of the BNS remnant, gravity plays a role and must be considered.}
%As it is well known, special relativity is not compatible with the inclusion of gravity, unless we are in the Newtonian regime of a nearly flat spacetime and low velocities. Given a classical gravitational acceleration $\vec{g} = - \vec{\nabla}\phi$, the source terms to use in the RMHD equations can be derived from the Newtonian limit of the so-called $3+1$ formulation of the general relativistic (GRMHD) equations (see e.g. \citealt{ECHO_2007, Rezzolla_Zanotti_2013}). In particular, the \textit{lapse function} for a diagonal metric (e.g. Schwarzschild with gravity due to a central mass $M$) is $\alpha = \sqrt{-g_{tt}}\simeq 1 +\phi$, with $\phi = - GM/r$ and $|\phi | \ll 1$, usually approximated as $\alpha\simeq 1$, unless when spatially derived in the source terms on the right-hand side of the conservative system. When the rest mass term dominates the total energy and momentum densities, one may then write
%\begin{equation}
%     \vec{f} = - {\cal E}\, \vec{\nabla}\alpha \simeq D \,\vec{g} \,, \qquad 
%     {\cal P} = - \vec{m}\cdot \vec{\nabla}\alpha \simeq D\gamma\vec{v}\cdot \vec{g},
%\end{equation}
%and for low velocities compared to the speed of light the classical MHD limit is recovered ($\gamma\to 1$, $D\to\rho$). We remind once again that this is just a simple and not rigorous approach, given that the curvature of spacetime and the contribution by other sources of energy are neglected, only the use of a full GRMHD code would provide a correct solution in more extreme regimes.
Following \citet{Pavan_etal_2021} and \citet{Pavan_etal_2023}, gravity is treated like an external force in the RMHD equations, with the corresponding acceleration $\vec{g}$ which is not simply provided by the standard Newtonian term due to the central mass $M_0/M_\odot = 2.596$ of the remnant, but it also takes into account the (outward) radial pressure gradients caused by hot and magnetized material during the BH formation, a term that is gradually decreasing in time. As done in the cited works, we model this scenario by providing an effective mass and then we use
\begin{equation}
    \vec{g}= - \frac{G [M_0 - M_\mathrm{eff}(r,t)]}{r^2}\hat{\vec{r}} .
\end{equation}
\change{The explicit value of $M_\mathrm{eff}$ is computed by first recovering its value at $r = r_{\mathrm min}$ and the collapse time $t = t_c\;{\mathrm s}$ by solving the angle-averaged radial momentum equilibrium, i.e., yielding}
\begin{equation}
M_\mathrm{eff}(r_{\mathrm min},t_c) = -\left[r^2\DS\frac{1}{\overline{\rho}G}\left(\DS\der{\overline{p}}{r} + \DS\frac{1}{8\pi} \der{\overline{B^2}}{r} - \DS\frac{\overline{(\vec{B}\cdot\nabla)B_r}}{4\pi}\right)\right]_{r_{\mathrm min}, 0}.
\end{equation}
\change{Consequently, the radial and temporal dependence is provided by:}
\begin{equation}
M_\mathrm{eff} = M_0 - \max\left(0.1 - \DS\frac{r - r_\mathrm{min}}{r_G - r_\mathrm{min}}\right)M_\mathrm{eff}(r_\mathrm{min},t_c)\,\mathrm{e}^{-\DS\frac{t + \tau_j}{\tau}},
\end{equation}
\change{
where $r_G \sim 1033.65\,\mathrm{km}$ is the gravity characteristic radius, $\tau_j = -t_c = 30 \,\mathrm{ms}$ represents the delay time between the neutron stars collapse and the jet launch, and 
}
\begin{equation}
\tau = \tau_j + (\tau_d - \tau_j)\sin^2{\theta}
\end{equation}
\change{
represents the gravitational timescale\footnote{\change{a typo is present in \citet{Pavan_etal_2021, Pavan_etal_2023}, although this affects only the manuscript and not the corresponding code.}} with $\tau_j = 0.3 \,\mathrm{s}$ the accretion time-scale of the black hole-disk system.
Note that in all the simulations we have adopted the point-mass approximation for Newtonian gravity.
}

%--------------------------------------------------------------------
\section{Axisymmetric parameter study}
\label{Sec:results2D}
%--------------------------------------------------------------------

To understand the impact of the resistivity we performed a set of different axisymmetric 2D simulations, thus reducing the necessary computational cost.
Despite the issues described in the literature (e.g., \citealt{Lopez-Camara_etal_2013, Harrison_etal_2018}) related to the propagation of the jet along the axis at $\theta = 0$, the relatively short computational time required to evolve the jet up to $r\sim90000\km$ allowed us to perform a parameter study.
To properly disentangle the impact of the magnetic dissipation on the jet propagation and dynamics, we chose 4 configurations, two with Lundquist numbers (defined as $S = v_aL/\eta$ where $v_a$ is the Alfv\'en speed and \change{$L = 150\km$ is the length scale}) $S = 10^2$ (acronym S2) and two with $S= 10^3$ (acronym S3).
For each chosen value of the Lundquist number, we have performed one simulation assuming $v_a = c$ (acronym C), i.e. constant resistivity in time and space, and one where we recover $v_a$ in its exact form (acronym V)
\begin{equation}
\label{eq::alfven}
    v_A/c = \max\left(\frac{|\vec{B}|}{\sqrt{\rho c^2 h + |\vec{B}|^2}},0.001\right)
\end{equation}
where the lower limit is set to avoid potential cells where $\eta = 0$, which may not be properly solved by the IMEX scheme \citep{Bugli_etal_2014, Ripperda_etal_2019a}.
We show, in the left panel of Fig. \ref{fig:rhoeta_2D}, the values of the resistivity at the final time of each simulation as a function of space, overlapped by the contour lines of the passive scalar tracer (dashed and dotted, respectively corresponding to the values of 0.1 and 0.01). 
As a preliminary consideration, we note that since within the jet, the Alfv\'en speed is close to the speed of light, cases with the same Lundquist number yield similar values of resistivity (only within the jet).
\change{Additionally, we have reported a schematic summary of the key parameters in Table \ref{tab:parameters}.}

\begin{table}
\centering
\caption{\footnotesize \change{Relevant parameters for the simulations discussed in the paper.}}
\begin{tabular}{lccccccc}
 \hline
 Case & $S$ & $v_a/c$ & $\eta\;[\mathrm{cm}^2/\mathrm{s}]$ & $t_{\mathrm br} [\s]$ & $r_{\mathrm br} [\km]$ \\
 \hline \\
 S2V & $10^2$ & Eq. \ref{eq::alfven} & 
 $5\times10^{12-15}$ & 0.177 & $20.7\times 10^3$ \\
 S3V & $10^3$ & Eq. \ref{eq::alfven} & 
 $5\times10^{11-14}$ & 0.182 & $21.4\times 10^3$\\
 S2C & $10^2$ & 1 & 
 $5\times10^{15}$ & 0.167 & $19.9\times 10^3$\\
 S3C & $10^3$ & 1 & 
 $5\times10^{14}$ & 0.177 & $20.9\times 10^3$\\
 3D  & $10^2$ & Eq. \ref{eq::alfven} & 
 $5\times10^{12-15}$ & - & - \\
 S4C & $10^4$ & 1 & 
 $5\times10^{13}$ & 0.206 & $23.1\times 10^3$ \\
  IDEAL & $\infty$ & 1 & 
 $0$ & 0.202 & $22.8\times 10^3$\\
 \hline \\
\end{tabular}
   \tablefoot{The columns, from left to right, show, respectively, the case name, the corresponding Lundquist number, the Alfv\'en speed formula for the resistivity, the minimum and maximum available resistivity values, and the time and radius of slow breakout (see the text in this section). The cases S4C and IDEAL are discussed only in the appendix.}
    \label{tab:parameters}
\end{table}

\change{
Within the single cases with variable resistivity, the jet is more diffusive when compared to the ambient medium, due to the higher values of the jet's local Alfv\'en speed.}
Therefore, due to the \change{impact of the} numerical resistivity \change{(see Appendix \ref{App:ideal}), the case S3V is dominated by the physical non-ideal processes only within the jet (e.g. where the Alfven speed is closer to the speed of light) or at shorter distances, while the numerical one dominates only beyond the interface between the jet and the ambient medium at larger distances due to the lower resolution caused by the stretched grid (as shown next subsection and in Appendix \ref{App:ideal}).}
Conversely, the dynamics of the case S2V \change{is} still dominated by the physical resistivity also outside the jet and through the entire radial domain extension.
We point out that, due to the magnetization of the post-merger wind, resistive processes can also occur at larger distances from the injection site and in the post-merger environment.

\begin{figure*}
  \centering
  \includegraphics[width=0.97\textwidth]{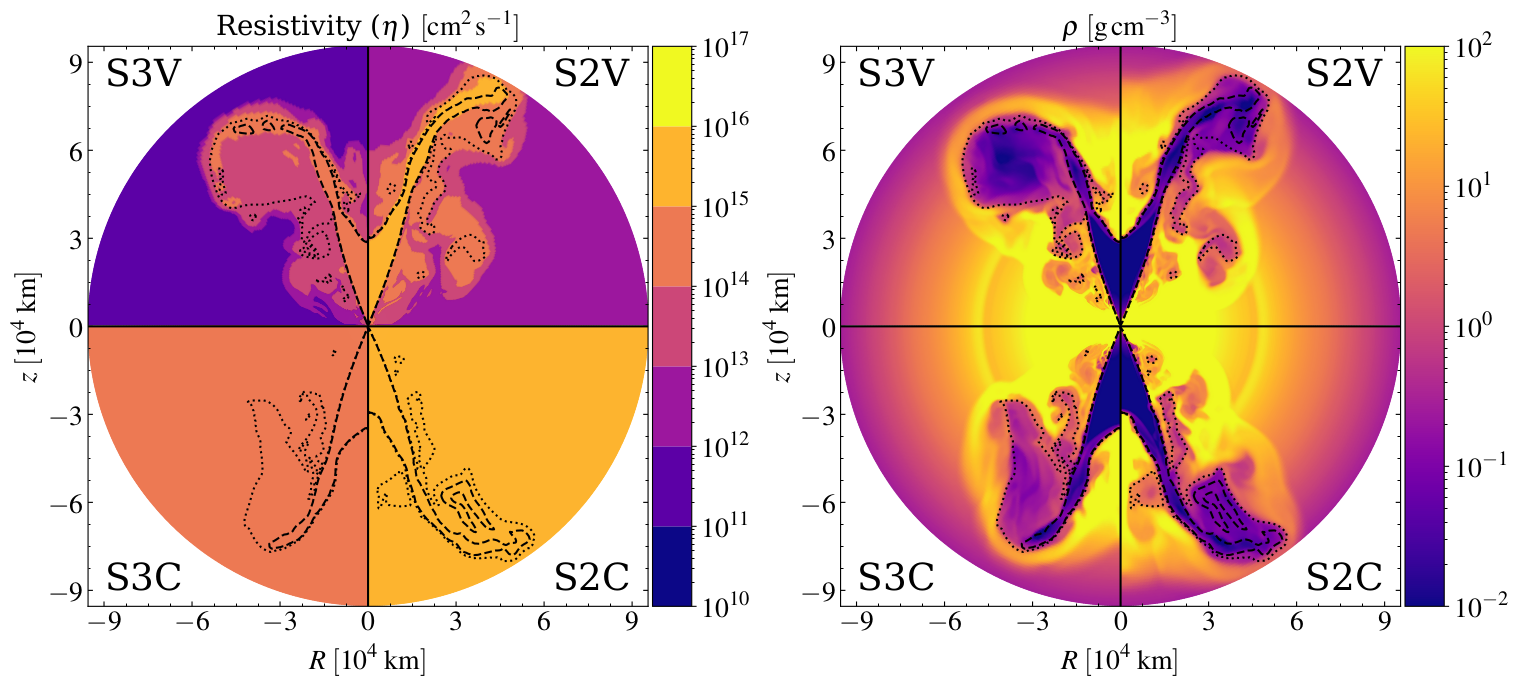}
  \caption{\footnotesize Resistivity $\eta$ (left panel) and rest-mass density $\rho$ (right panel) at the final time of the simulation ($t\sim0.5\s$) for the different resistivity cases. The dashed black line represents the tracer contour line at level 1\%, while the solid lines only aim at delimiting the different cases. Note that in this and all the 2D plots, the angle $\theta$ is between $0\degree$ and $90\degree$ and the negative cylindrical coordinates are simply a post-processing transformation to enhance the comparison quality.}
  \label{fig:rhoeta_2D}
\end{figure*}

The density at the final time of the simulations is shown in the right panel of Fig. \ref{fig:rhoeta_2D}.
Firstly, we notice a ring-like structure, which corresponds to the "slow" waves traveling in the expanding wind (the quotes are adopted since the resistivity affects the wave propagation, as shown in \citealt{Mignone_etal_2018}), which is expanding into the unmagnetized atmosphere.
\change{
Due to the expansion of the magnetized environment into the atmosphere, different propagating waves yield a series of shocks and discontinuities in the external regions, which propagate at different characteristic speeds.
Since the propagation of the environment is, although slower than the jet, still relativistic, at the final time of the simulation the jet does not break into the unmagnetized atmosphere but travels into different environment regions characterized by the propagation of the slow waves.
This propagation of the environment leads to an accumulation of material and a magnetic field, which affects the jet propagation and dynamics.
Due to the different behavior between these two propagation stages, we have reported the time of the slow breakout in Table \ref{tab:parameters}. The impact of the propagation of the jet into the strongly magnetized (inner) environment region and the weakly magnetized (outer) environment region is discussed in the next subsections of this paper.

Moreover, the magnetic resistivity affects the shape and the position (see the next subsection) of the termination jet lobes and the jet backflow, as well as the shape and number of the recollimation regions present (although we point out that these regions cannot be associated with the \change{$m=0$} plasma columns instabilities since, due to the axisymmetric assumption, the jet shapes like a funneled column subjected to the higher pressure from the ambient medium from both inside and outside the jet).
In particular, we notice that a constant resistivity is associated with a more uniform dissipation of the magnetic field close to the lobes, resulting in less (maximum 2 in each C case) recollimation regions.
Conversely, a variable (in time and space) resistivity leads to a more fragmented outer jet (3 and 5 recollimation zones can be detected, respectively, in the cases S2V and S3V).
While in these simulations the jet shape is constrained by the assumption of axisymmetry, we expect such features to impact the jet dynamics in fully 3D simulations strongly (see Section \ref{Sec:results3D}), both on the jet dynamics and the radiation signature associated (e.g. \citealt{Dong_etal_2020}).}

\begin{figure}
  \centering
  \includegraphics[width=0.47\textwidth]{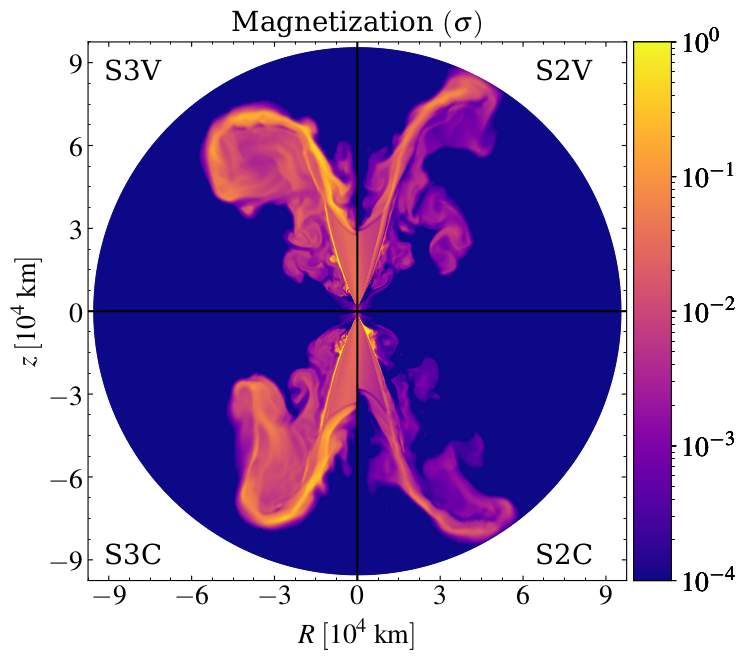}
  \caption{\footnotesize \change{Magnetization $\sigma$ at the final time of the simulation ($t\sim0.5\s$) for the different resistivity cases.}}
  \label{fig:magnetization}
\end{figure}

\change{
Finally, in Fig. \ref{fig:magnetization} we show the magnetization (defined as in Eq. \ref{eq::sigma}).
Due to the dissipation caused by the resistivity, all the cases show a lower magnetization at larger distances and much lower values than the magnetization injected at $t = 0$.
Moreover, the resistivity affects the jet magnetization both at the jet axis and the termination regions, yielding higher values of $\sigma$ for lower values of $\eta$.
In particular, following the criteria of \citet{Bromberg_Tchekhovskoy_2016}, the S3 cases show a moderate jet magnetization ($\sigma\lesssim1$) even at larger distances, while the S2 cases can be considered as weakly magnetized (i.e. $\sigma \ll 1$, more similar to the jets investigated in \citealt{Gottlieb_etal_2020}).
A comparison with an ideal run is done in Appendix \ref{App:ideal}, while a more quantitative analysis of the magnetic energy dissipation is performed in the following Section.
}

%-----------------------------------------------------
\subsection{Jet energy and dynamics}

As shown in \citet{Mattia_etal_2023}, the magnetic resistivity is expected to play a role in terms of the \change{shape and dynamics} of potential current sheets, energy conversion (from magnetic to thermal), and production of unaligned electric field (from the ideal $-\vec{v}\times\vec{B}$ direction) and dissipated power.
\change{Due to the lower grid resolution in the outer domain regions, the development of small-scale (i.e. few $\mathrm{km}$) turbulent structures is strongly suppressed for $t\gtrsim0.1\s$ in our simulations.
Nevertheless, the physical diffusion of the electromagnetic field remains dominant over the numerical one (see Appendix \ref{App:ideal} for a comparison with an ideal run where the electric field is imposed to be $\vec{E} = -\vec{v}\times\vec{B} $).}

In the top and bottom left panels of Fig. \ref{fig:jetdin_2D} the evolution of the jet electromagnetic energy, \change{with respect to} the case S3C at the final time and the non-electromagnetic energy (i.e. the thermal + kinetic without the rest-mass contribution, see, e.g., \citealt{Ricci_etal_2024})
\begin{equation}
\begin{array}{lcl}
    E_{\rm int}  & = & \DS\int\xi\left[\rho\gamma^2(h - 1)c^2 - p\right]r^2\sin\theta \rm{d}r\rm{d}\theta \rm{d}\phi \\ \noalign{\medskip}
    E_{\rm mag}  & = & \DS\int\xi\left[\DS\frac{E^2 + B^2}{2}\right]
    r^2\sin\theta \rm{d}r\rm{d}\theta \rm{d}\phi \\ \noalign{\medskip}
    E_{\rm kin} & = & \DS\int\xi\left[D(\gamma - 1)c^2\right]r^2\sin\theta \rm{d}r\rm{d}\theta \rm{d}\phi
\end{array}
\end{equation}
is shown.

\begin{figure*}
  \centering
  \includegraphics[width=0.97\textwidth]{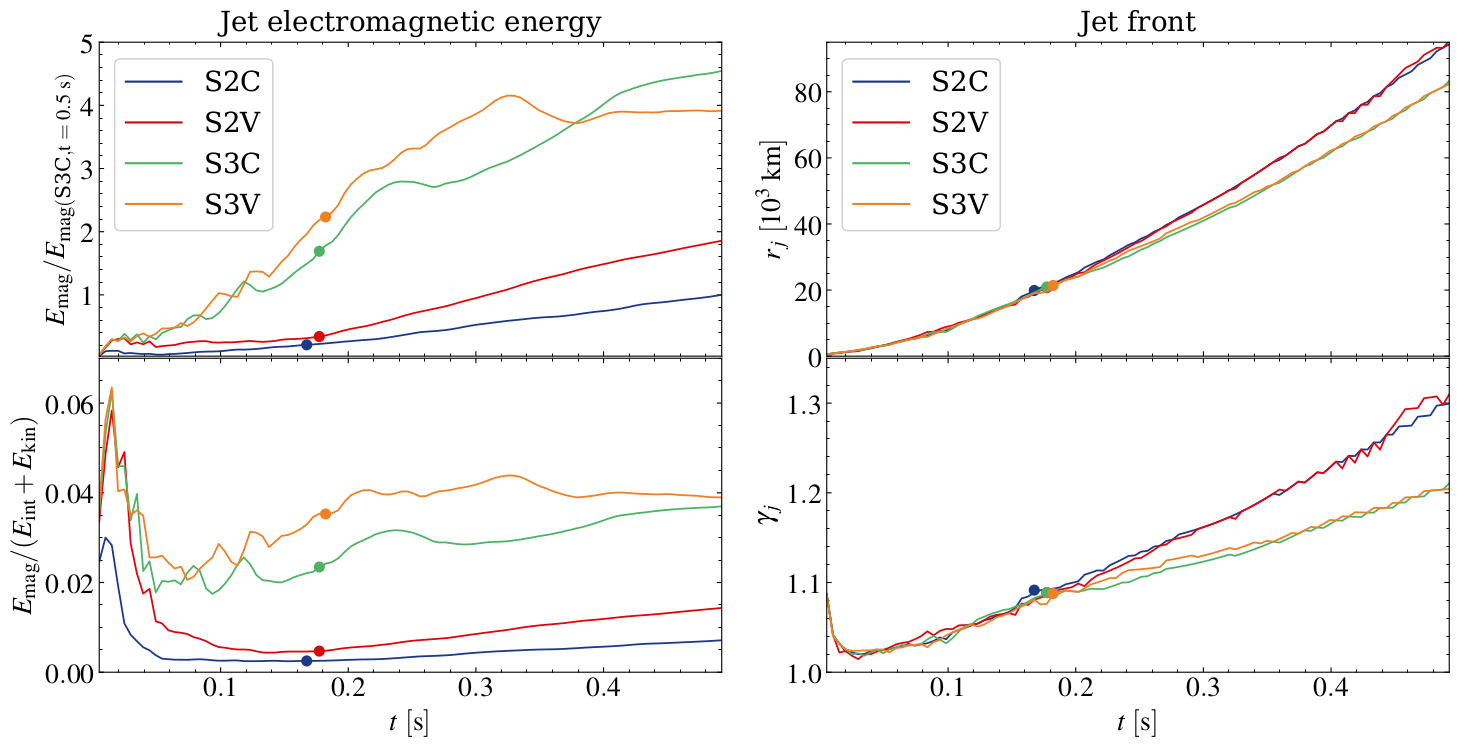}
  \caption{\footnotesize Time evolution of the jet energetics and dynamics. 
  The electromagnetic energy contribution (compared to the one of the case S3C at the final time in the top left panel and to the kinetic + internal energy in the bottom left panel) and the jet front (position in the top panel, and Lorentz factor in the bottom panel) are shown for different 2D runs. \change{The circle dot represents the time of slow breakout.}}
  \label{fig:jetdin_2D}
\end{figure*}

As a first consideration, we notice that the S2 cases, \change{in the top-left panel of Fig. \ref{fig:jetdin_2D}}, show a quasi-constant behavior until $t\sim0.2\s$ and (only the S2C case) a decrease around $t\sim0.05\s$.
Such a lack of increase is the result of the strong diffusivity processes that balance the continuous energy injection at the inner boundary.
However, at later times (i.e. when the jet has pierced the remnant and starts propagating through the \change{weakly magnetized environment region}) the diffusivity plays a much less crucial role, due to the lack of the magnetic field \change{in such region}, leading to an increase in the electromagnetic energy in every case during the later evolutionary stages (despite its decay at the injection site described by Eq. \ref{eq:decay_vars}).
On the other hand, the S3 cases show an increase in electromagnetic energy up to $t\sim0.4\s$, with the S3C case showing such an increase throughout the entire evolution.
As for the previous case, two distinct phases can be detected: in the early stage of the simulation (i.e. $t\lesssim0.1\s$) the dissipative processes play a crucial role in the interaction between the magnetized wind and the jet; once the jet has passed the \change{slow breakout} radius, the magnetic resistivity becomes less relevant since the \change{external environment region is very weakly magnetized} and the injection process becomes dominant.
The reason behind the "plateau" visible at the end of the simulation S3V may be related to a shift in the dominant diffusive mechanism, since at larger distances the numerical resistivity may overcome the physical one in this particular case.
Therefore, an uncontrolled numerical resistivity may slow even further the \change{increase} of the magnetic energy \change{related to the jet injection}.

In the bottom left panel of Fig. \ref{fig:jetdin_2D} the impact of electromagnetic energy with respect to the kinetic and thermal jet energy is shown.
Note that the jet remains hydrodynamically (thermally) dominated since the electromagnetic contribution remains below $7\%$ for all the simulations and settles around $4\%$ and $\lesssim2\%$ respectively for the S3 cases and the S2 cases.
In the very initial evolutionary stages, the electromagnetic contribution to the jet's total energy increases (concerning both its initial value and the non-electromagnetic contribution, as shown also in the top left panel); however, this increase of the electromagnetic energy fraction is reverted already at $t\sim0.015\s$, due to the combination of the interaction between the jet and the remnant medium and the magnetic dissipation caused by the magnetic resistivity.

All the cases (but S2C) show a similar peak in the electromagnetic energy impact, while the suppression of the magnetic energy contribution (which is reverted around $t \sim 0.1\s$ for the S3 cases and $t \sim 0.2\s$ for the S2V case) yields a different minimum before a subsequent increase due to the propagation of the jet in the \change{weakly magnetized environment region}.
At this point, due to the lack of a magnetic field, the jet propagation proceeds unaltered (i.e. no strong turbulence is present outside the outer lobes). As a result, the only physical processes able to impact the electromagnetic field are magnetic dissipation and jet injection, leading to a slow increase in the contribution of electromagnetic energy.
However, as stated previously, due to the lack of magnetization, electromagnetic field injection becomes the dominant process, especially for low Lundquist numbers.

We point out that these simulations \change{cannot reveal} a heating process by solely looking at the energy output, since energy conversion can also occur outside the resistive regime due to the simple interaction between the jet and the surrounding environment.
However, this is a clear dissipation process triggered by the magnetic resistivity which affects the impact of the electromagnetic component of the jet energy up to $\sim30\%$ on the jet energy balance.
Nevertheless, as pointed out at the beginning of the section, the jet propagation and dynamics are also affected by the magnetic resistivity.
\change{Due to the energy balance resulting from the different simulations, one would expect that a more dissipative jet (because of the energy conversion) would propagate on different timescales through the external medium.
If on one hand, the dissipation of the magnetic field quenches the magnetic acceleration, on the other hand, the heating dissipative processes are able to increase the heat reservoir converted to kinetic energy during the jet expansion.}
In order to assess more quantitatively the impact of the magnetic dissipation, we computed (see the top right panel of Fig. \ref{fig:jetdin_2D}) the jet front as the maximum radial distance from the core at which the Lorentz factor is above 2 (note that the remnant has an initial maximum Lorentz factor of $\sim 1.22$, well below our constraint).
\change{The position of the jet front is also used to compute the time at which the slow breakout occurs, i.e. when the jet front has pierced the discontinuity associated with the environment's slow wave propagation.
The time and position of the slow breakout are reported in Table \ref{tab:parameters}.}

%in the two following ways:
%
%\begin{itemize}
%    \item as the maximum radial distance from the core at which the jet-to-ambient ratio (defined by the scalar tracer) is above $1\%$;

%    \item as the maximum radial distance from the core at which the Lorentz factor is above 2 (note that the remnant has an initial maximum Lorentz factor of $\sim 1.22$, well below our constraint).
    
%\end{itemize}
%
%Note that the two methods show significant differences only at the early evolutionary stages, providing a better agreement with time (the relative difference is $\lesssim10\%$ at $t\sim0.25\s$ and $\lesssim5\%$ at $t\sim0.4\s$ with decreasing trend). 
%Such differences during the initial evolutionary stages should not be surprising due to the strong initial interaction between the jet and the ambient medium.
As expected, once the jet has reached the \change{weakly magnetized environment region}, the resistivity impacts the jet propagation by $\approx10000\km$ (which corresponds to approximately the $11\%$ of the distance covered by the jet at $t\sim0.5\s$).
We notice that the jet front spatial propagation is not linear, suggesting a non-trivial temporal evolution of its propagation.
By looking at the Lorentz factor associated with the jet front propagation, shown in the bottom right panel of Fig. \ref{fig:jetdin_2D}, we can divide the jet propagation into three different stages:

\begin{itemize}

\item initial stage ($t < 0.05\s$), the jet is still confined within the (more dense) post-merger environment, the jet front Lorentz factor decreases since the matter is slowed down by both the accumulation of matter at the jet axis and the turbulence caused by the interaction between the jet and the external medium in all our simulations

\item \change{slow breakout} stage ($0.05\s < t < 0.2\s$), the jet breaks through the post-merger environment and starts propagating through the \change{weakly magnetized environment region}, all the simulations show a similar increase of the Lorentz factor

\item final stage ($t > 0.2\s$), the jet propagates through the low-density and low-pressure \change{weakly magnetized environment region}, keeping the same internal structure, the resistivity strongly affects this stage since the magnetic resistivity is the only magnetic process that can impact the jet energy.
    
\end{itemize}

\change{Note that the positions and times of the slow breakout show very little or no difference between the cases if compared with the evolution at the later stages.}
Such division differs partially from the one obtained by looking at the electromagnetic energy because changes in the electromagnetic energy do not immediately reflect on their hydrodynamical counterpart (and most of the Ohmic dissipation takes place in the inner radii, see next subsection).
Nevertheless, the three evolutionary stages detected in the bottom left panel of Fig. \ref{fig:jetdin_2D} are consistent with the different trends in jet propagation.

The different trend of the Lorentz factor leads to a difference of about $\sim7\%$ between the S2 and the S3 cases at the final time of our simulation.
Since this relative difference in the Lorentz factor increases with time (at $t = 0.1$ and $t = 0.3$ the differences are, respectively, around $\sim1\%$ and $3\%$), we expect the dynamics of jets characterized by different resistivity models, to show differences also at larger times.

%-----------------------------------------------------
\subsection{Parallel field and dissipated power}

\begin{figure*}
  \centering
  \includegraphics[width=0.97\textwidth]{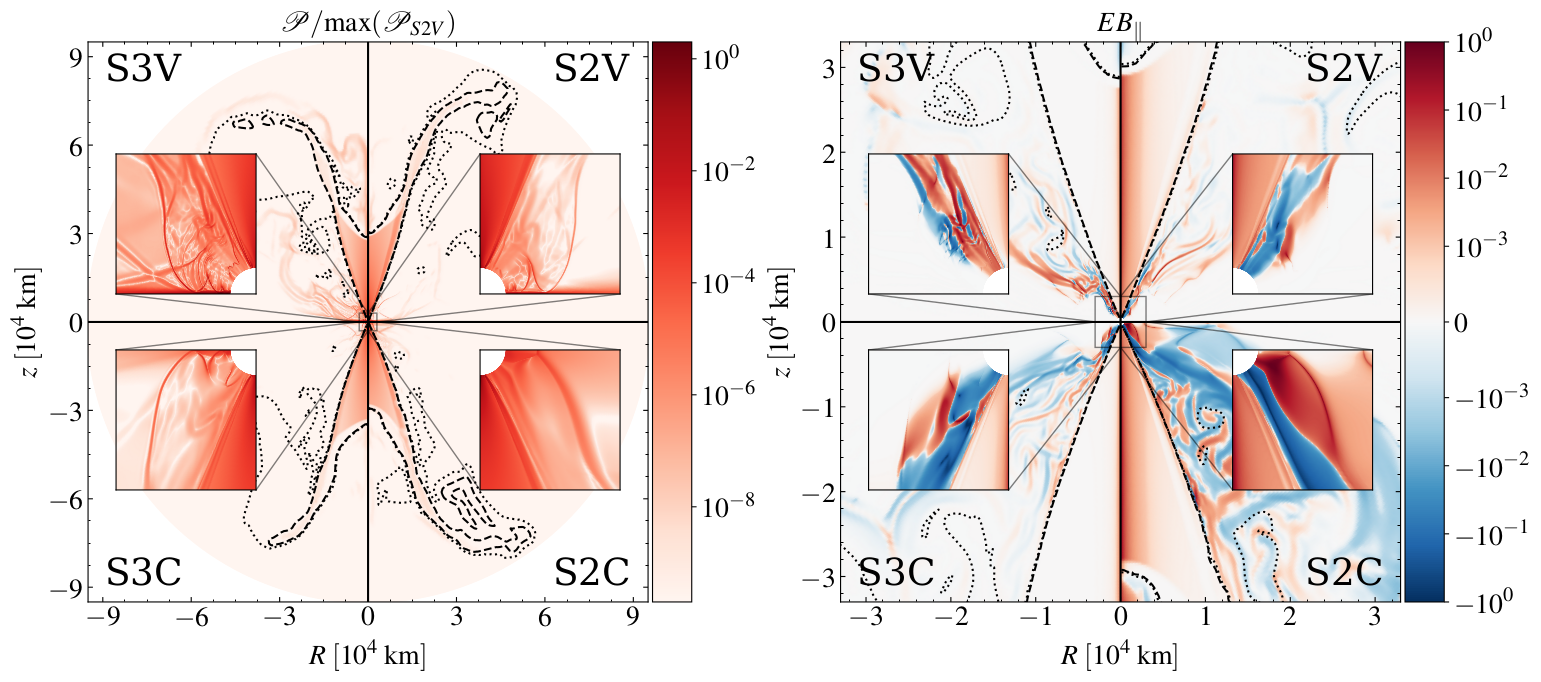}
  \caption{\footnotesize Dissipated power (left panel) and normalized parallel component of the electric magnetic field (right panel) for the different resistivity cases at $t\sim0.5\s$. The zoomed regions show the inner region up to $3000\km$ in both $R$ and $z$ directions for both panels.}
  \label{fig:ebpower_2D}
\end{figure*}

The magnetic resistivity is also responsible for the generation and amplification of electric field which deviates from the $-\vec{v}\times\vec{B}$ assumption \citep{DelZanna_etal_2016}, as shown in Fig. \ref{fig:ebpower_2D} as dissipated power (left panel, normalized to the maximum value obtained from the S2V case at the end of the simulation) and parallel electromagnetic field (right panel), computed, respectively, as
(see also Eq. \ref{eq:emu+bmu})
\begin{equation}
\begin{array}{lcl}
{\cal P} & =& \eta j^2 = e^2/\eta = \gamma^2\DS\frac{(\vec{E} + \vec{v}\times\vec{B})^2 - (\vec{v}\cdot\vec{E})^2}{\eta} \\ 
\noalign{\medskip}
EB_\parallel & = & \DS\frac{\vec{E}\cdot\vec{B}}{\sqrt{E^2B^2} + \epsilon}
\end{array}
\end{equation}
where $\epsilon$ serves the purpose of avoiding undefined divisions.

In agreement with \citet{Mattia_etal_2023}, a stronger resistivity yields a higher dissipated power in the cases where the resistivity is constant in time and space.
In all cases, most of the power is dissipated within the innermost region and slowly decays as the jet propagates until the termination region (i.e. $\sim3\times10^4\km$, where the ${\cal P}$ is several orders of magnitude below the innermost one).
However, the power is not dissipated only within the jet, since for all the cases a non-negligible power is dissipated in the remnant ambient region.
In particular, a filamentary structure arising from the equator ($\theta = 90\degree$) is present.
The structure of such filaments is much more complex in the variable resistivity cases, where multiple gradients are present. In contrast, in the constant resistivity cases, the structure outside the jet shows less turbulence and a much smoother behavior.
This is not surprising since narrow regions of high local Alfv\'en speed, outside the jet, may yield a higher level of dissipated power that propagates outwards, leading to weakly resolved features whose spatial size corresponds to few numerical cells and depends on the local magnetic resistivity.
Moreover, the S3V case shows significant dissipated power outside the jet and also at greater distances from the inner boundary (near both the equator and the jet axis) and a higher level of dissipated power when compared to the S3C case. Such a feature is not present in any of the other cases, and its nature seems to be numerical (similar issues in the resistive electric field were found in the low resistivity cases of \citealt{Mattia_etal_2023}).

By looking at the right panel of Fig. \ref{fig:ebpower_2D}, we notice that the S2C case yields a much more diffused "parallel field region" than the other 3 cases, especially close to the midplane region and in the \change{weakly magnetized environment region}.
Such a feature is a direct consequence of the strong magnetic resistivity through the entire domain, therefore not only within the jet but also in the magnetized ambient medium.
As expected, the S2 cases show a stronger parallel component close to the jet axis, extending to $\sim30000\km$ (the maximum jet position along the axis $\theta = 0$).
Conversely, the S3 cases do not lead to a strong non-ideal electric field ($\gtrsim1\%$ of its magnitude) within the jet and show deviations from the ideal regime only from the interface between the jet and the ambient medium, where the Kelvin-Helmholtz instability and magnetic reconnection are likely to take place and generate and amplify resistive electric field.

All the cases (but S3V, likely due to the low resistivity outside the jet) show a clear double parallel field inversion: within the jet, the resistive electric field features a parallel component, which becomes antiparallel outside the interface between the jet and the ambient.
Additionally, another resistive field reversal, with different magnitudes, is present in the inner \change{weakly magnetized environment regions}.
Such inversion of the resistive electric field is clearly enhanced in the presence of a stronger physical resistivity. and is strictly correlated to a velocity inversion (radial between the jet and remnant, toroidal within the remnant region, see the next section)
At the early evolutionary stages, such inversion can trigger the development of the Kelvin-Helmholtz Instability (KHI, \citealt{Chow_etal_2023}) which is a favorable environment for the amplification of non-ideal magnetic field and magnetic dissipation and reconnection \citep{Sironi_etal_2021}.
Conversely, the S3V case shows no inversion at the jet-ambient interface, although a strong gradient in the parallel electromagnetic field magnitude is still present.

%-----------------------------------------------------
\subsection{Initial evolution}

\begin{figure*}
  \centering
  \includegraphics[width=0.97\textwidth]{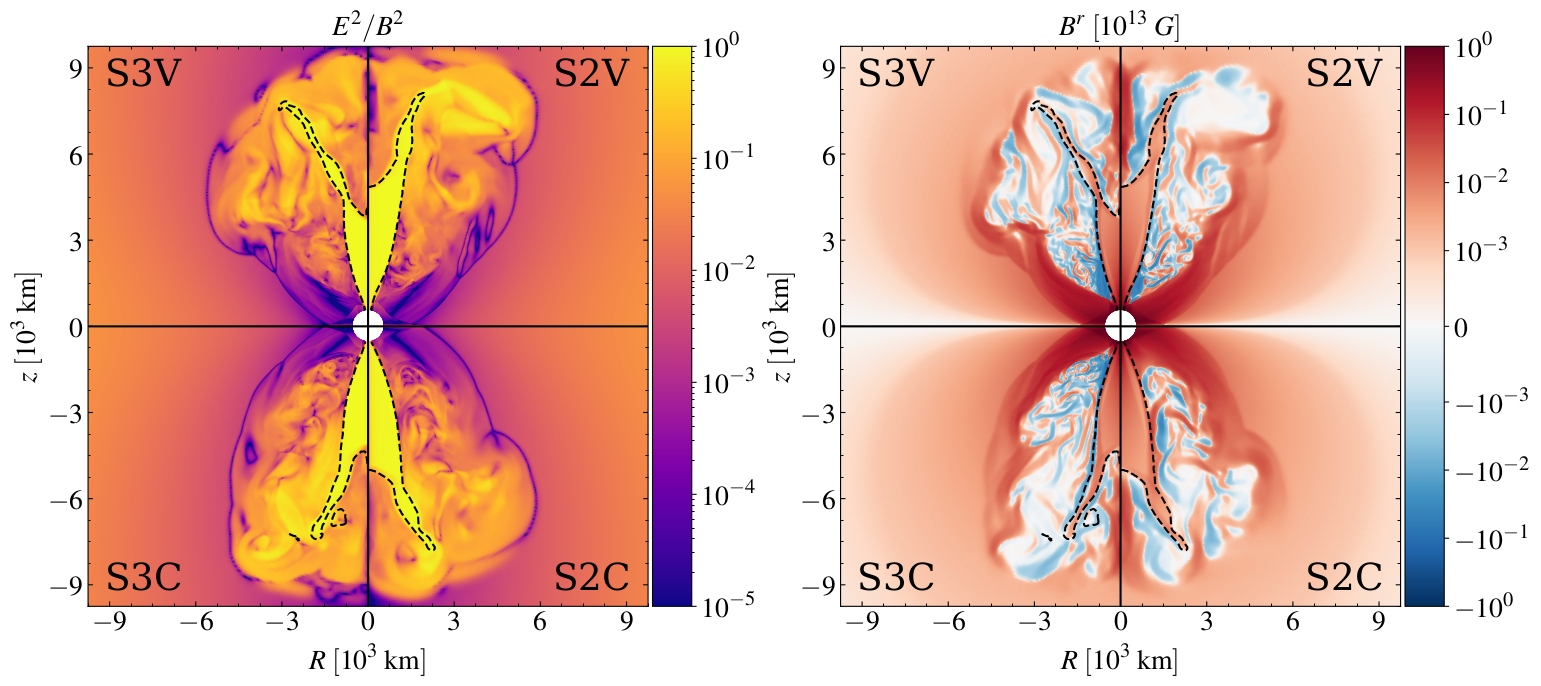}
  \caption{\footnotesize Electromagnetic field properties at the early simulation stages. The ratio between the electric and magnetic energy (left panel) and radial component of the magnetic field (right panel) are shown, for the different resistivity cases, at $t = 0.1\s$, superimposed by the contour lines of the scalar tracer at 0.1.}
  \label{fig:ebearly_2D}
\end{figure*}

To provide a more meaningful investigation of short GRB jets, we need to combine the scales at which the resistive processes take place with the large propagation scale reached by the jet after the surrounding environment is pierced.
Although the lower resolution, at large distances, smears out much of the turbulence present within the jet, we can still investigate the impact of the resistivity more in detail when looking at the jet at early times (e.g. $t\sim0.1\s$, which corresponds to the $20\%$ of the final time of our simulations).
In particular, the electromagnetic features of the resistive simulations can provide a clear indication of magnetic reconnection and particle acceleration sites.
The ratio between the electric and magnetic field is reported in the left panel of Fig. \ref{fig:ebearly_2D}, superimposed by the contour lines of the passive tracer at level 0.1.

Here the jet structure shows, already at this evolutionary stage, significant differences in shape, both inside and outside the regions where the jet has interacted with the post-merger environment.
%The high resistivity cases show, as expected, a higher degree of collimation (i.e. a smaller $R$ is reached for equal values of $z$) and a faster propagation close to the jet axis.
The interrelation between the jet magnetic field and its dynamics has been extensively studied in a more general context \citep{Leismann_etal_2005, Moya-Torregrosa_etal_2021} and in more specific jet types, such as AGN jets \citep{Ricci_etal_2024} and GRB-jets \citep{Gottlieb_etal_2020, Pavan_etal_2023}
The region affected by the interaction between the jet and the external medium shows a high level of turbulence and a ratio $E^2/B^2$ between $1\%$ and $10\%$ (as shown in the left panel of Fig. \ref{fig:ebearly_2D}), suggesting that magnetic reconnection (which is likely to occur in such location due to the KHI and the plasma column instabilities) may play a crucial role in particle acceleration (as already shown in \citealt{Beniamini_Giannios_2017}) in GRBs as well.

Reversals in the radial component of the magnetic field are also an indicator of potential magnetic reconnection sites. 
Due to the injected magnetic field structure, we expect the magnetic field to play a significant role in the jet structure and collimation.
The presence and the shape of magnetic field reversal zones, where the magnetic field reaches a minimum due to a change of polarity, are strongly affected by the resistivity adopted \citep{Ripperda_etal_2019b, Mattia_etal_2023}.
In particular, since the formation of thin magnetic layers with different polarity takes place where the jet and the post-merger environment have strongly interacted, the physical resistivity must overcome the numerical one so that the reconnecting region does not depend on the resolution and the numerical algorithms adopted.
This is the case for the S3V model, where such regions are not resolved and their width is, for the most part, limited to very few computational cells.
Conversely, the cases S3C and S2V show regions favorable for magnetic reconnection associated with wider magnetic layers.
In addition, in the turbulent ambient region (where the Alfv\'en speed is of the order of $\sim10\%$ of the speed of light), we find that the resistivity associated with the reconnecting regions is comparable for the cases S3C and S2V (as in the left panel of Fig. \ref{fig:rhoeta_2D}), while within the jet (where the Alfv\'en speed approaches $c$) the cases with the same Lundquist number also show a similar resistivity 
Lastly, the S2C case resistivity is too strong to favor the formation of turbulent reconnecting regions, yielding a more diffused structure similar to the one found in the more diffusive cases of \citet{Mattia_etal_2023}.

%--------------------------------------------------------------------
\section{Full 3D simulation}
\label{Sec:results3D}
%--------------------------------------------------------------------

\begin{figure*}
  \centering
  \includegraphics[width=0.97\textwidth]{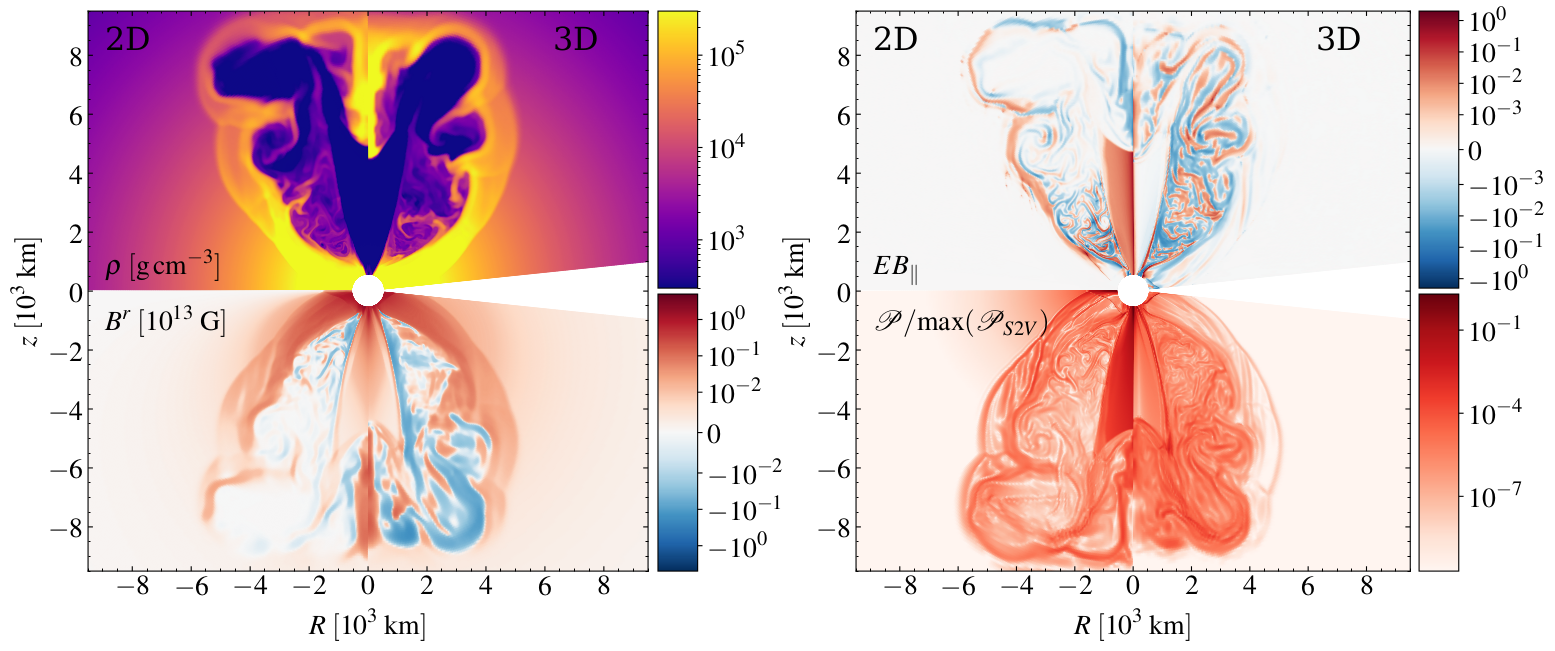}
  \caption{\footnotesize Comparison with axisymmetric and fully 3D simulations. Density (top part) and radial magnetic field (bottom part) are shown in the left panel, while normalized parallel electromagnetic field (top part) and jet-dissipated power (bottom part) are shown in the right panel. The left part of each panel represents the case S2V, while the right part represents the cut of the 3D simulation at $\phi = \pi/2$ and $\theta \geq \pi/2$.}
  \label{fig:rhoeb_3D}
\end{figure*}

% The assumption of axisymmetry (with respect to the jet propagation axis) is known to influence the jet evolution in the dense surrounding environment.
% As shown in different works \citep[e.g.,][]{Lopez-Camara_etal_2013,Harrison_etal_2018}, 
% 2D axisymmetric simulations are plagued by the unphysical accumulation of dense material on the jet propagation axis, which slows down the jet itself by deforming its head into a plume shape.

% As shown in the fully 3D numerical simulations of \citet{Lopez-Camara_etal_2013, Bromberg_Tchekhovskoy_2016} (although the latter refers to core-collapse GRB jets), the jet breaks the assumption of axisymmetry already at the early stages of the propagation.
As shown in fully 3D numerical simulations of relativistic jets \citep[e.g.][]{Mignone_etal_2010_jet, Lopez-Camara_etal_2013, Bromberg_Tchekhovskoy_2016}, the assumption of axisymmetry can be broken already at the early stages of jet propagation.
% The onset of the $m = 1$ kink instability \citep{Mignone_etal_2010_jet, Gottlieb_etal_2021} and the absence of axisymmetric boundary conditions prevent the unphysical accumulation of dense material along the jet injection axis after the jet has pierced through the magnetized environment \citep{Lopez-Camara_etal_2013, Harrison_etal_2018}, which slows down the jet itself by deforming its head into a plume shape.
The absence of axisymmetric boundary conditions, in particular, prevent the unphysical accumulation of dense material along the jet injection axis,  which can alter the propagation itself leading to unrealistic configurations of the jet angular structure \citep[see, e.g., Figure~1 in][]{Mignone_etal_2010_jet}.
Moreover, the turbulent interaction (which we expect to be strongly affected by the resistive and dissipative processes considered in this paper) between the jet and its surrounding environment can fully develop only in 3D.
For these reasons, we performed a fully 3D simulation with analytical initial and boundary conditions (as for the 2D axisymmetric simulations) until $t = 0.1\s$, which is approximately the time at which the jet started piercing the post-merger environment and reached a distance of $r\sim10^4\km$ from the central object.
% Until this phase, the plume shape can still be present due to the interaction between the jet and the heavier environment.

\begin{figure*}
  \centering
  \includegraphics[width=0.97\textwidth]{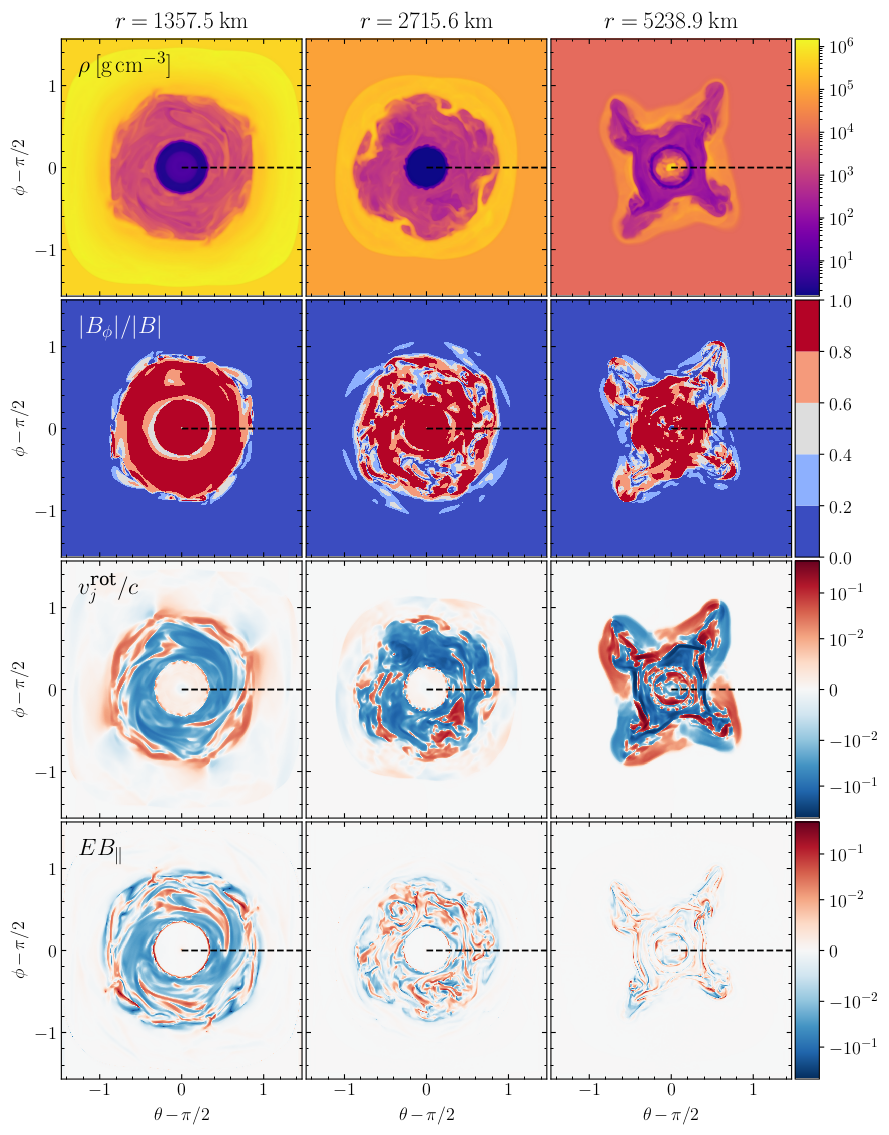}
  \caption{\footnotesize Jet angular profiles at different radii. From left to right, the distance from the center is, respectively, $1357.5\km$, $2715.6\km$ and $5238.9\km$. The rest-mass density, toroidal over total magnetic field, jet rotation velocity, and normalized parallel electromagnetic field are shown from top to bottom. The dashed black lines represents the cut at $\phi = \pi/2$ and $\theta \geq \pi/2$, as in Fig. \ref{fig:rhoeb_3D}.}
  \label{fig:shape_3D}
\end{figure*}

In Fig. \ref{fig:rhoeb_3D}, we compare a 3D simulation cut (at $\phi = \pi/2$ and $\theta \geq\pi/2$) with the corresponding 2D axisymmetric case.\footnote{Note that, especially at large radii, the angular structure of the jet may strongly vary; therefore, variations for different angles must be taken into account in the comparison itself (see Fig. \ref{fig:shape_3D}, \ref{fig:full_3D} and the relative discussion at the end of this section).}
On the left panel, we show the density (top part) and the radial magnetic field (bottom part), while on the right panel, the parallel electromagnetic field (top part) and the dissipated power (bottom part) are shown. 
Here the radial and scalar components are unaffected by the setup rotation, so the comparison does not require any coordinate transformation.
Note that the accumulation of material closer to the jet axis here present is not related to numerical instabilities, but to the injection of a lighter jet into a more massive environment.
Due to the onset of the KHI, which takes place already in the merger \citep{Kiuchi_etal_2015} and the earlier injection phase and represents a powerful magnetic field amplification mechanism, the 3D case shows a higher level of turbulence at the layers between the jet and its surroundings. In particular, a stronger radial magnetic field with steeper gradients is present, as shown by the multiple field reversals.
A stronger magnetic field can be associated with more efficient magnetic reconnection and particle acceleration \citep{Sironi_Spitkovsky_2014}.

\begin{figure*}
  \centering
  \includegraphics[width=0.47\textwidth]{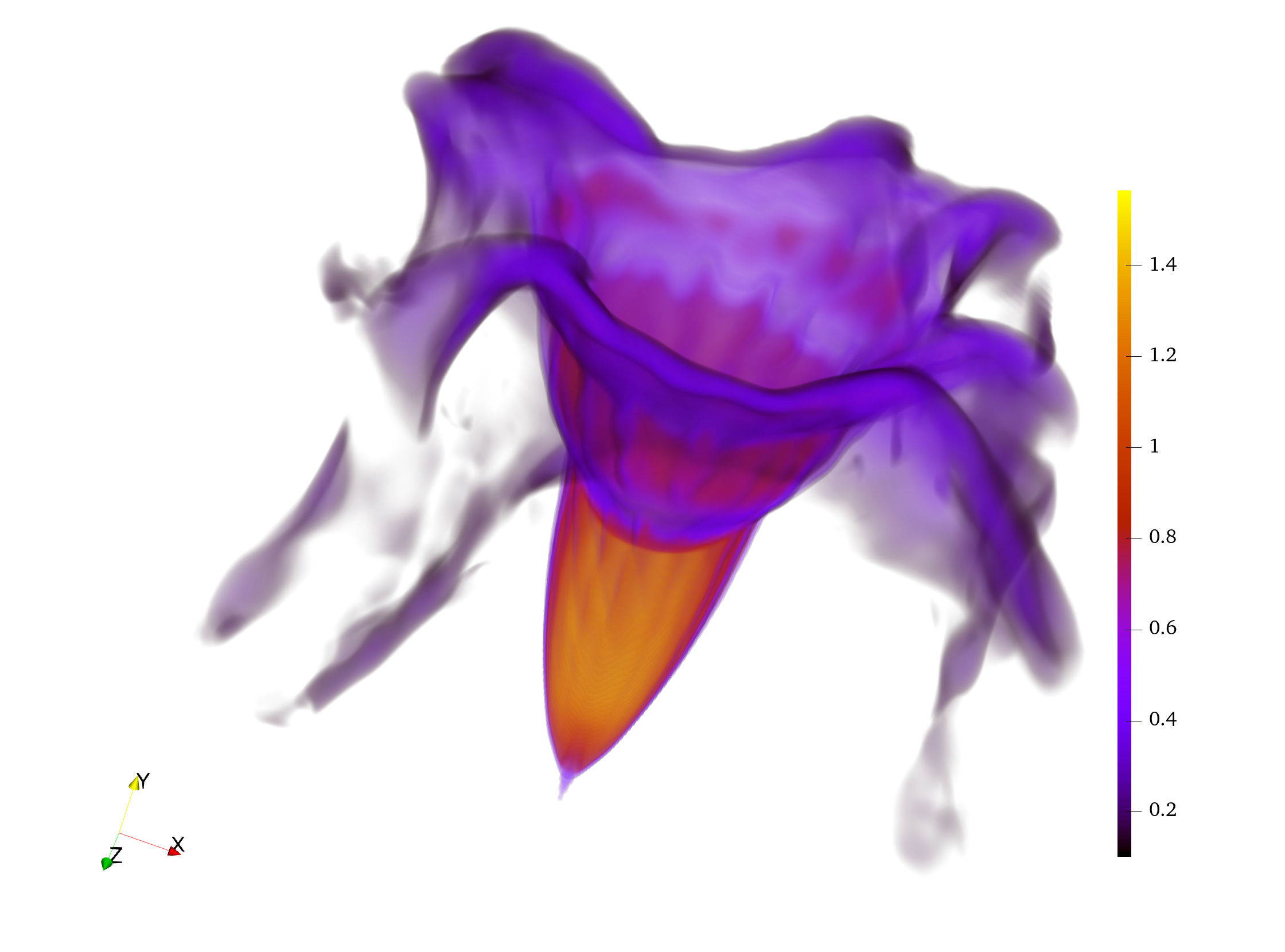}%
  \includegraphics[width=0.47\textwidth]{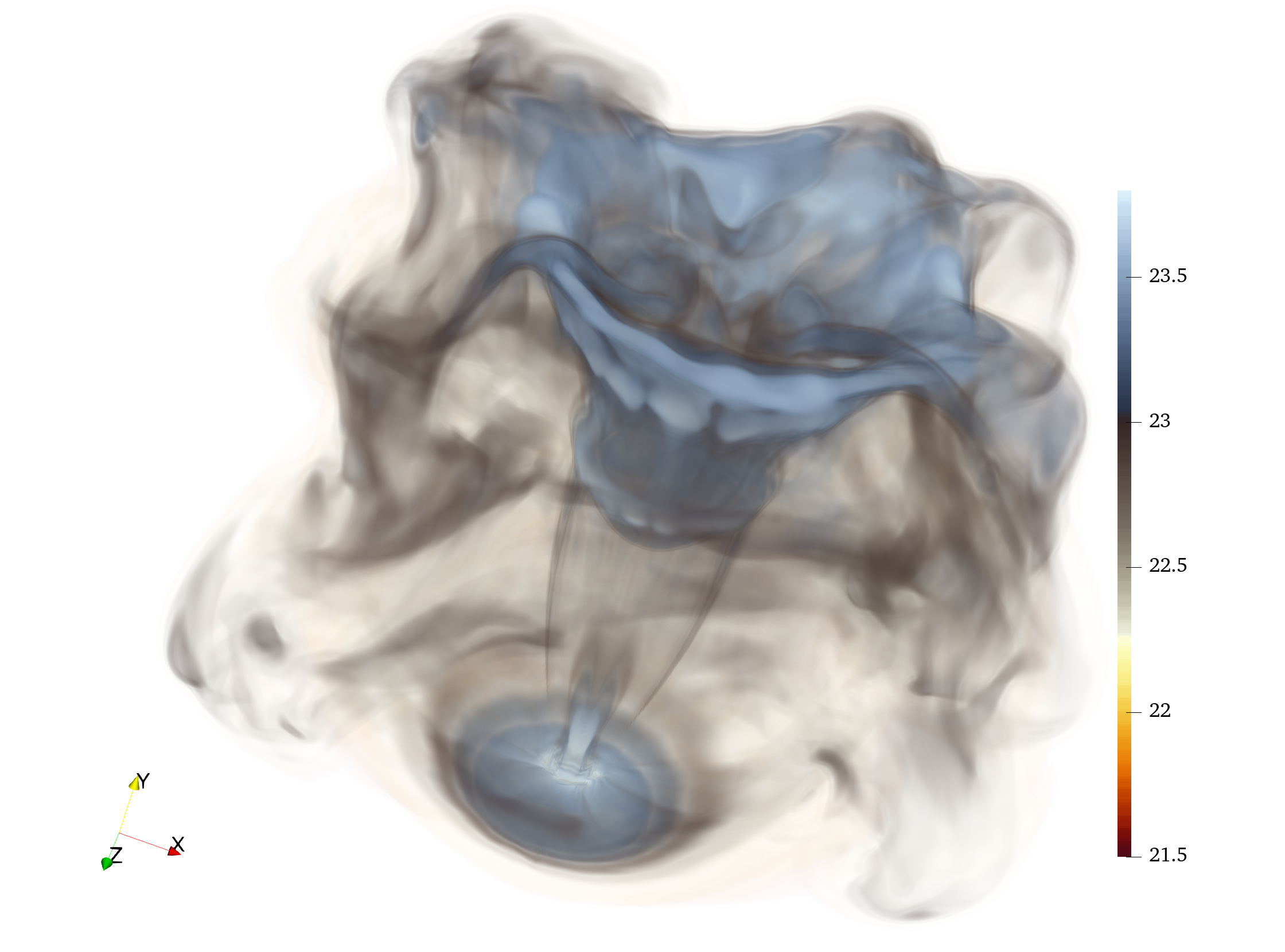}
  \caption{\footnotesize Fully 3D jet structure. The Lorentz factor's logarithm and magnetic energy density's (in $\mathrm{G}^2$ units) logarithm are displayed, respectively, on the left and right panels.}
  \label{fig:full_3D}
\end{figure*}

Due to the enhanced turbulence, the 3D simulation shows a different resistive electromagnetic field (Fig. \ref{fig:rhoeb_3D}, right panel). 
More specifically, closer to the axis the electric field is quasi-perpendicular to the magnetic field (the parallel component is below $0.01\%$), while outside the jet the turbulent structure is, as for the previous variables here considered, more prominent.
Here charged particles can also be accelerated due to the non-perpendicularity of the electromagnetic field, provided that their gyroradius is much smaller than the variations in the electromagnetic field.
\change{ Since the minimum jet magnetic field is approximately $B_\mathrm{min}\approx 10^{11}\mathrm{G}$, the maximum gyroradius for a charged particle traveling at the speed of light would be
   $3.3\times10^{-4} {\mathrm m}$ for a $\mathrm{TeV} $ electron/proton, with a linear dependence of the gyroradius on the particle energy.}

In Fig. \ref{fig:shape_3D}, we report the \change{azimuthal} distribution at different radii of several fluid variables, to assess more deeply the level of axisymmetry present in this simulation (which, as described in the previous section, starts with an axisymmetric jet launched into an axisymmetric environment).
All the right panels, which correspond to a radial distance of $5238.9\km$ from the center, show an instability with mode $m = 4$. %that leads to the filamentary structure of the left panel of Fig \ref{fig:full_3D} (where the logarithm of the Lorentz factor is shown, while on the right panel, the logarithm of the magnetic energy is displayed).
Such a feature should be related to the Rayleigh-Taylor (RT) instability, triggered by the accelerated jet expansion in the denser environment, which drives mixing between the two fluids \citep[][]{Mignone_etal_2010_jet, Gottlieb_etal_2021}.
For this to occur, the magnetic tension due to the toroidal component of the magnetic field should not be the dominant one (in the presence, as in our case, of a magnetized jet-environment).
Here, the ratio between the toroidal component (retrieved by applying the inverse transformation of Eq. \ref{eq:vec_tilt}) and total magnetic field, shown in the second row of Fig.~\ref{fig:shape_3D}, can help to understand the origin of the instability itself better.
At lower radii (left panel), two regions where the magnetic field is toroidally dominated are detected: within the jet and the surrounding cocoon; such zones are separated by a thin poloidally dominated layer. At larger radii, the toroidal component is no longer always dominant in the second region, and the RT instability is free to develop there, leading to a strong mixing of material with the external medium. 

%\emph{The recollimation of the filaments seen in Fig \ref{fig:rhoeta_2D} becomes (see the left panel of Fig. \ref{fig:full_3D}) a $m = 0$ plasma column instability (unlike in the axisymmetric case where a cylindrical slab was pinched).
%On the other hand, the density at lower distances shows a higher level of axisymmetry, as shown also in the Lorentz factor, radial magnetic field, and magnetic energy.
%In fact, we notice that the 3D turbulence is fully developed only at $r > 2.5\times10^3\km$, as shown in the middle and right columns of Fig \ref{fig:shape_3D}.}

We also computed the jet rotation, shown in the third row of Fig.~\ref{fig:shape_3D}.
A double inversion in the jet rotation is visible here: the inner rotation change at the jet boundary and an additional flip in the rotation direction at larger angles.
Note that such a trend is recovered also in the 2D simulations, where there is no equivalent behavior of the magnetic field.
At larger distances (middle column) the jet rotation becomes more ``patchy'' with a general counterclockwise predominance and some localized regions with clockwise movement.
Once the jet has approached its peak (right column), the $m=4$ instability has fully taken place, and the backflow of material 
%(also seen in Fig. \ref{fig:full_3D})
is divided into four regions, each with two opposite rotation directions.
The presence of such strong gradients in shear velocities cannot exclude a contribution to the enhancement of the turbulence level due to KHI, which also in the relativistic regime, as in classical MHD, is affected by the relative importance of magnetic field components, aligned or transverse to the shear \citep{Bucciantini_DelZanna_2006}.

Finally, the parallel electromagnetic field $EB_\parallel$ is shown in the last row of Fig.~\ref{fig:shape_3D}.
Here, the symmetry with respect to the jet injection axis is broken already at $r = 1357.5\km$, due to the higher level of turbulence outside the jet, as seen in Fig.~\ref{fig:rhoeb_3D}.
%and the right panel of Fig \ref{fig:full_3D})
At larger distances, the parallel electromagnetic field becomes weaker due to the magnetic field diffusion.
Nevertheless, few regions where particles can undergo non-thermal acceleration are present even at $r = 5238.9\km$.
This indicates that resistivity becomes even more relevant in fully 3D simulations due to their enhanced turbulence (compared with 2D simulations).

To conclude, Fig. \ref{fig:full_3D} shows the 3D distributions of the Lorentz factor (left panel) and magnetic energy density (right panel), in logarithmic scale, in the lab frame at the end of the simulation.
A clear anticorrelation between the two quantities can be noted.
In particular, a stronger magnetic field is present closer to the jet front, while at the inner propagation region (i.e. where the jet has maximum velocity) the magnetic energy density has lower values.
As shown in \cite{Pavan_etal_2023} and our axisymmetric simulations, we expect the magnetic energy to be converted into kinetic energy after the jet has pierced the environment and broken out into the \change{weakly magnetized environment region}.
Moreover, where the magnetic energy has not been converted into kinetic energy, the jet shows a stronger accumulation of the former at the boundaries between the jet and the environment, while the core remains weakly magnetized. 
Such hollow core is also found in different simulations (e.g. \citealt{Nathanail_etal_2021}).
Here, its origin is associated with the lateral expansion of the jet due to the interaction with the denser environment since the jet has not broken into the less dense atmosphere.

By looking at the hydrodynamic (left panel of Fig. \ref{fig:full_3D}) and the electromagnetic (right panel) jet components, we notice a similar shape in the presence of enough jet material (which is correlated with a high Lorentz factor). However, due to the environment's magnetic field, the turbulent interaction outside the jet cone yields a more turbulent structure and extends much further in the electromagnetic component, suggesting that reconnection and dissipation processes can more efficiently take place in such a scenario.
The enhanced turbulent level, coupled with a more complex behavior of the resistive quantities, is expected to be enhanced even more by injecting the jet into a non-axisymmetric realistic environment.

%-------------------------------------------------------------------
\section{Conclusions}
\label{Sec:conclusions}
%-------------------------------------------------------------------

In this paper, we have presented the first resistive relativistic simulations of short GRB jets.
By combining a suite of high-order numerical algorithms with a two-component radial grid, we were able to properly capture the dissipative effects occurring at the early propagation stages, i.e. when the jet has not pierced the magnetized remnant yet.
Our results can be summarized as follows:

\begin{enumerate}

\item We performed four different 2D axisymmetric simulations up to $t\sim0.5\s$ with varying prescriptions of resistivity (i.e. $ = 10^2$ or $S = 10^3$ and uniform or variable resistivity), allowing us to disentangle the resistive processes during the jet propagation;

\item A higher resistivity is related to a higher degree of electromagnetic dissipation, yielding a more thermal jet that propagates faster through the remnant and the \change{weakly magnetized environment region}. While at the earlier propagation stages, the resistivity plays a key role mostly in the electromagnetic jet components, once the jet has pierced the magnetized remnant environment, the resistivity becomes crucial not only in the diffusion of the electromagnetic field but also in the jet front position and velocity;

\item Most resistive effects, i.e. the generation and amplification of parallel electromagnetic field and dissipated power, are sited in the innermost jet region. However, a filamentary structure is visible closer to the equator whose gradients strongly depend on the resistivity profile. The parallel electromagnetic field shows a double inversion in most cases with a tight interrelation between a lower diffusivity and steeper gradients;

\item Already at the early evolutionary stages, magnetic resistivity strongly affects the electromagnetic jet counterpart.
In particular, a lower resistivity yields a higher level of turbulence and plasma instabilities at the interface between the jet and the ambient medium, which coupled with the strong electric field, makes them an up-and-coming site for the acceleration of non-thermal particles. Conversely, a higher resistivity is associated with a smoother magnetic field whose configuration is less favorable for the onset of magnetic reconnection;

\item We performed a fully 3D simulation up to $t\sim0.1\s$ to assess the deviation from axisymmetry even in a fully axisymmetric initial configuration, as well as the features related to the absence of boundaries at the jet axis.
A consequence of the loss of axisymmetry is the lack of accumulation of material at the jet axis and the amplification of the magnetic turbulence farther from it, including a more turbulent resistive electric field outside the jet;

\item At larger distances from the injection sites, the jet undergoes a plasma instability of mode 4, due to the Rayleigh-Taylor instability, which affects its shape and generates a filamentary backflow that undergoes a series of recollimation regions.  We also found that the resistive components of the electromagnetic field are more susceptible to the breaking of the axisymmetry, strongly deviating from it already closer to the jet base, indicating that fully 3D simulations will augment the impact of the resistivity on the jet evolution.

\end{enumerate}

In conclusion, the importance of physical resistivity, which is not dependent on the grid resolution and the suite of numerical algorithms adopted (provided that it is higher than the numerical one) is clearly stated, as well as the importance of fully 3D simulations.
Due to the complexity of the physical processes involved here, which should encompass very different lengths and timescales, the grid resolution can strongly impact the outcome in the presence of only an uncontrolled numerical resistivity. 
Conversely, adopting a physically motivated, consistent resistivity model makes it possible to reproduce the observational features found in nature more accurately.

In a companion paper (Paper II), we will explore the impact of the resistivity on jet propagation in a more realistic non-axisymmetric environment, which will be imported consistently from BNS numerical relativity simulations as in \citet{Pavan_etal_2021, Pavan_etal_2023, Pavan_etal_2024}.

%---------------------------------------------------
\begin{acknowledgements}
We thank the anonymous referee for the valuable comments and suggestions that greatly improved the quality of the manuscript.
GM acknowledges D. Crocco for the assistance and valuable comments on the Python routines. 
LDZ acknowledges support from the ICSC—Centro Nazionale di Ricerca in High-Performance Computing; Big Data and Quantum Computing, funded by European Union - NextGenerationEU.
AP and RC acknowledge support from the European Union under NextGenerationEU, via the PRIN 2022 Project ``EMERGE: Neutron star mergers and the origin of short gamma-ray bursts'', Prot. n. 2022KX2Z3B (CUP C53D23001150006), and further support by the Italian Ministry of Foreign Affairs and International Cooperation (MAECI), grant number US23GR08.  
All the simulations were performed on GALILEO100 at CINECA (Bologna, Italy).
In particular, we acknowledge CINECA for the availability of high performance computing resources and support through awards under the ISCRA initiative (Grant IsB27\_BALJET) and through a CINECA–INFN agreement (providing the allocation INF24\_teongrav). 
\end{acknowledgements}

%-------------------------------------------------------------------
\section*{Data availability}
%-------------------------------------------------------------------

The PLUTO code is publicly available. Simulation data will be shared at a reasonable request by the corresponding author.

%-----------------------------------------------------------------
\bibliographystyle{aa.bst}
\bibliography{paper}

%-----------------------------------------------------------------
\appendix

%-------------------------------------------------------------------
\section{\change{Comparison with} the ideal run}
\label{App:ideal}
%-------------------------------------------------------------------

To assess the contribution of the physical resistivity, as opposed to the numerical one, we performed two additional runs with $S = 10^4$ (S4V and S4C) and a run by solving the ideal equations with the same numerical methods adopted for the resistive simulations (except for the time integration, which is now an explicit $3^{\rm rd}$-order Runge-Kutta method).
Note that, despite the same general structure, the FORCE Riemann solver differs from its resistive version since the computation of the fastest wave propagation, as well as the fluxes, is not the same.

\begin{figure}
  \centering
  \includegraphics[width=0.47\textwidth]{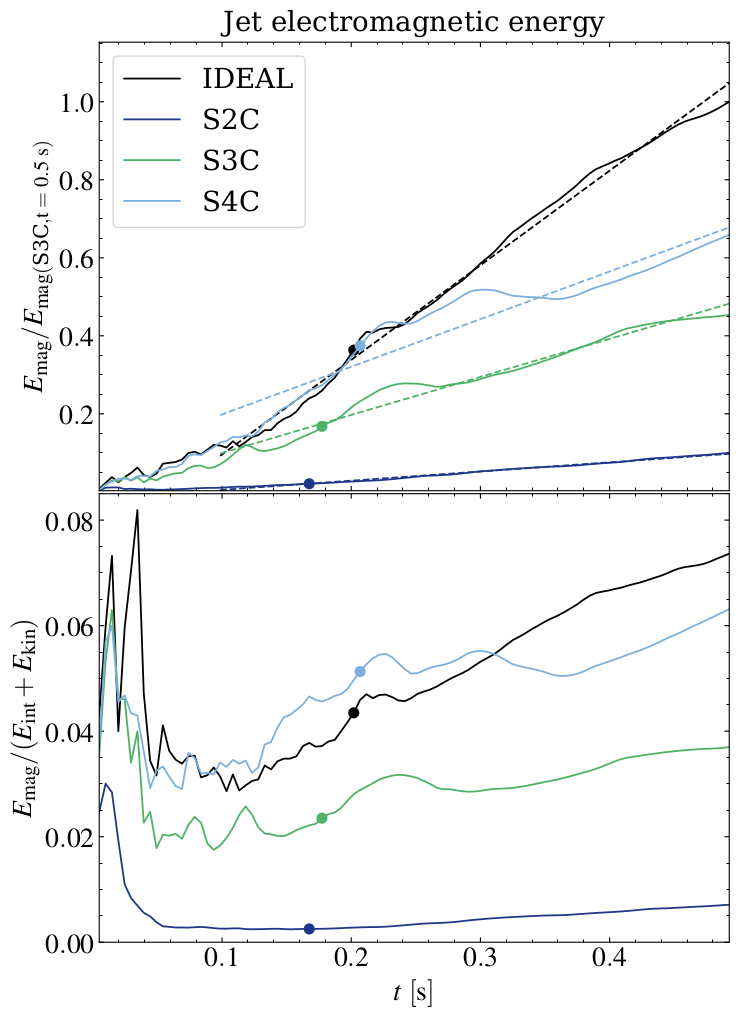}
  \caption{\footnotesize electromagnetic energy contribution (compared to the one of the case S3C at the final time in the top panel and to the kinetic + internal energy in the bottom panel) for different values of the Lundquist number and constant resistivity (colored lines) and ideal run (black line). The dashed lines are the linear approximation of the evolution of electromagnetic energy at later times. The circle dot represents the slow breakout.}
  \label{fig:Energy_app}
\end{figure}

\subsection{Energy budget and magnetic turbulence}

The numerical resistivity, which is intrinsically present in every simulation, may differ, especially at larger distances from the merger where the grid is less refined due to the stretching.
Such difference is clearly visible in Fig. \ref{fig:Energy_app}, where the magnetic energy is reported as a function of time (similarly to Fig. \ref{fig:jetdin_2D}).
Here we compared only cases with constant resistivity (in time and space) to neglect the impact of the jet dynamics on the resulting resistivity.
Although these simulations are less realistic due to the simplified magnetic resistivity model adopted, they provide a better insight into the impact of the numerical diffusion.
All the cases but the S4C one show a very good agreement with a linear dependence, provided that the jet has pierced the post-merger environment (i.e. after $t\sim0.1\s)$. 
Due to the grid stretching, we expect the numerical resistivity to increase at larger radii, eventually overcoming the physical one for very high values of the Lundquist number $S$.
This seems to be the case of the S4C run, where a strong trend change can be detected around $t\sim0.25$.
Note that the energy increase of the case S4C and S3V (top left panel of Fig \ref{fig:jetdin_2D}), despite both being dominated by the numerical diffusion at large distances, show a very different 
behavior, very likely because of the interaction between the numerical and physical resistivity.
Note that, before $t\sim0.25\s$, the ideal and S4C cases almost overlap, indicating that the contributions of the numerical and physical (only in the S4C simulation) are comparable.
Moreover, due to the unpredictable nature of the numerical resistivity, the ideal case slope may strongly change with different grid resolutions or more/less accurate numerical algorithms.

The impact of the resistivity on the role of the jet's electromagnetic energy (compared to the thermal and kinetic energy) is shown in the bottom panel of Fig \ref{fig:Energy_app}.
Here a higher Lundquist number is related to more (although the jet is dominated by the thermal energy in all simulations) magnetized jets, due to the lower amount of diffusivity. 
As shown in the top panel, the ideal and S4C cases show a similar behavior up to $t\sim0.25\s$, where the latter case becomes more dominated by the numerical resistivity.
\change{Note that the slow breakout occurs at slightly later times in the S4C and ideal cases (see also Table \ref{tab:parameters}), although no additional statement can be made due to the intricate interrelation between the jet and the environment.}

\begin{figure}
  \centering
  \includegraphics[width=0.47\textwidth]{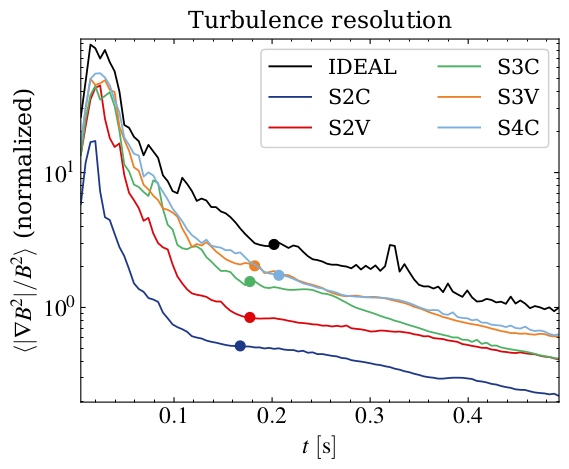}
  \caption{\footnotesize Effective turbulence resolution of the magnetic energy for different resistivity and ideal cases. The circle dot represents the slow breakout.}
  \label{fig:Turbulence_app}
\end{figure}

\begin{figure*}
  \centering
  \includegraphics[width=0.97\textwidth]{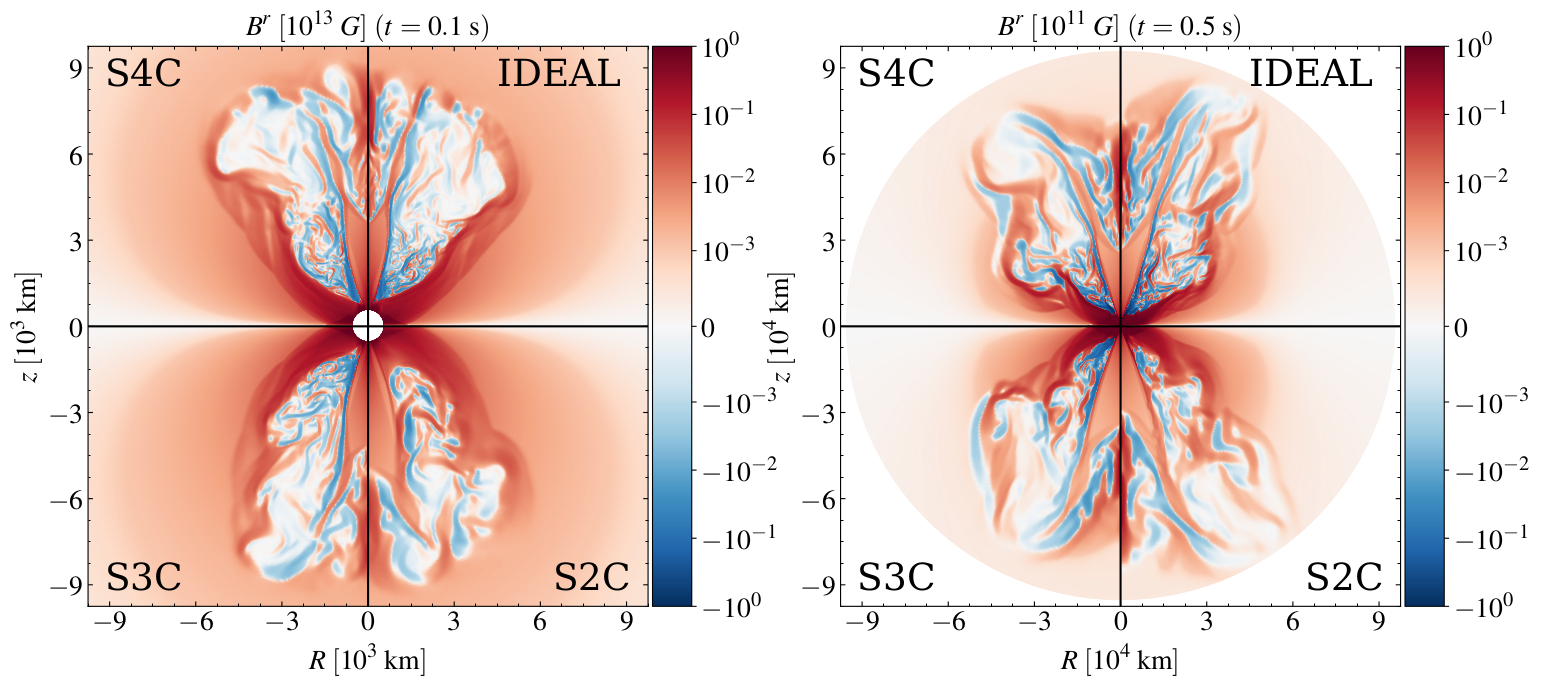}
  \caption{\footnotesize Electromagnetic field properties at the early and late simulation stages. The radial component of the magnetic field is shown, for the different constant resistivity and ideal cases, at $t = 0.1\s$ (left panel) and $t = 0.1\s$ (right panel).}
  \label{fig:magfield_app}
\end{figure*}

\change{
As in \citet{Mattia_etal_2023}, we have also checked the impact of the resistivity on the turbulence resolution by computing the normalized gradient $<|\nabla B^2|/B^2>$ for different runs with both constant and variable resistivity and the ideal run.
As shown in Fig. \ref{fig:Turbulence_app}, the turbulence level between the ideal and the resistive cases differs by a factor $\gtrsim2$.
We also note that the similarities between the cases S2V and S3C are also reflected here since the two cases show a similar level of turbulence at the end of the simulations.
The same statement holds for the cases S3V and S4C, which both seem to be affected by the numerical diffusion at larger distances.
From the energy budget, we have suggested that both S3V and S4C cases are affected by the numerical diffusion in the outer domain regions (where the resolution is the poorest); this hypothesis is strengthened by the very good agreement in terms of turbulence refinement that the two simulations reach.
Conversely, all the other simulations show a level of turbulence that is dictated by a physical, and therefore controllable, resistivity.}

\subsection{The electromagnetic field}

\change{
The main difference between the resistive and the ideal runs is the presence/absence of physical resistivity controlling the level of dissipation throughout the entire simulation.
The shape and size of current sheets or steep magnetic field gradients are strongly affected by the amount of dissipation present throughout the simulation.
Therefore, in Fig. \ref{fig:magfield_app}, we have compared the radial component of the magnetic field among different resistive simulations (with constant resistivity) and the corresponding ideal simulation.
The left and right panel show, respectively, the jet propagation at $t = 0.1\;\mathrm{s}$ and $t = 0.5\;\mathrm{s}$.
}

\change{
Both bottom panels, corresponding to the S2C and S3C simulations, show thicker (when compared with the other two cases) regions of the magnetic field with the same polarity, both at the early and later stages of the simulation. 
Nevertheless, the radial magnetic field shows a comparable magnitude in all the simulations, regardless of the employed resistivity.
Note that, in agreement with Fig. \ref{fig:Turbulence_app}, the S2C case shows the lowest turbulence resolution due to the high diffusion in both the jet and the ambient medium.
}

\change{
Conversely, the S4C and IDEAL cases show a more refined turbulence due to the lack of diffusion.
As expected, at lower distances from the center (left panel), the two cases yield a similar level of dissipation, confirming that the numerical diffusivity can be assessed to be around a Lundquist number of $S = 10^4$.
At larger distances and times (right panel), the case S4C suffers from higher numerical diffusion (compared to the ideal case) caused by the more diffusive Riemann solver and time-stepping algorithm. 
However, the two simulations remain very qualitatively similar and are affected by less dissipation than the cases with high resistivity, confirming that all the cases (but S3V) feature a higher physical resistivity (than the numerical one) at all the stages of the simulations.
}

\begin{figure}
  \centering
  \includegraphics[width=0.47\textwidth]{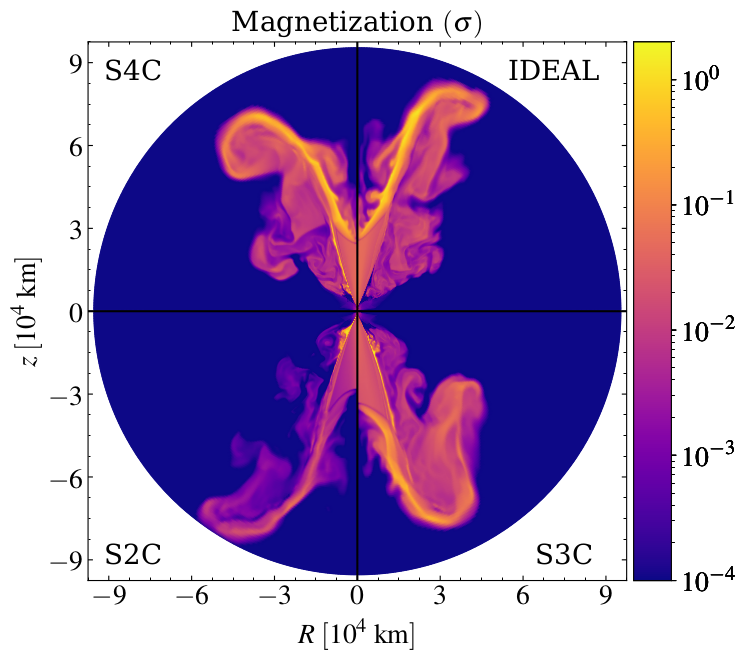}
  \caption{\footnotesize Magnetization $\sigma$ at the final time of the simulation ($t\sim0.5\s$) for the constant resistivity and ideal cases.}
  \label{fig:mag_app}
\end{figure}

\change{Finally, in Fig. \ref{fig:mag_app}, we have reported the spatial distribution of the magnetization $\sigma$, computed as in Eq. \ref{eq::sigma}.
In agreement with the previous subsection, we found that the highest magnetization levels are found in the S4C and ideal cases due to the weak diffusion of the magnetic field. 
In both cases, the jet is moderately magnetized at the final stages of the simulation, although in the termination stage (where the magnetic energy has not been fully converted to kinetic energy) the jet shows a stronger magnetization.
Note that the upper cases (S4C and IDEAL) show when compared to the lower cases (S2C and S3C) a stronger magnetization in both the injection, propagation, and outer phases due to the lack of resistivity.
}

\end{document}